\documentclass{aa}
\usepackage{graphicx}
\usepackage{natbib}
\usepackage{rotating}
\usepackage{bbm}

\bibpunct{(}{)}{;}{a}{}{,}
\newcommand{\be}{\begin{equation}}
\newcommand{\ee}{\end{equation}}
\newcommand{\vek}[1]{\mbox{\boldmath$#1$}}

\newcommand{\myarcsec}{\hbox{$.\!\!^{\prime\prime}$}}
\newcommand{\myarcmin}{\hbox{$.\!\!^{\prime}$}}
\newcommand{\dd}{\textrm{d}}

\begin{document}
   \title{GaBoDS: The Garching-Bonn Deep Survey}
   \subtitle{VI. Cosmic shear analysis\thanks{Based on observations made with ESO Telescopes at the La Silla Observatory}}

   \author{M.~Hetterscheidt\inst{1}
          \and P.~Simon\inst{1}
          \and M.~Schirmer\inst{1,2}
          \and H.~Hildebrandt\inst{1} 
          \and T.~Schrabback\inst{1}
          \and T.~Erben\inst{1}
	  \and P.~Schneider\inst{1} 
          }

   \offprints{Marco Hetterscheidt, e-mail: mhetter@astro.uni-bonn.de \\
$\,^\dagger\,$ Founded by merging of the Sternwarte, Radioastronomisches Institut and Institut f\"ur Astrophysik und Extraterrestrische Forschung der Universit\"at Bonn.}

   \institute{$^{1}$Argelander-Institut f\"ur Astronomie$^\dagger$, Universit\"at Bonn, Auf dem H\"ugel 71, 53121 Bonn, Germany\\
              $^{2}$Isaac Newton Group of Telescopes, Apartado de correos 321, 38700 Santa Cruz de La Palma, Tenerife, Spain}

   \date{Received 2006; accepted}

   \abstract
{}
{We present a cosmic shear analysis and data validation of 15 square degree high-quality $R$-band data of the Garching-Bonn Deep Survey obtained with the Wide Field Imager of the MPG/ESO 2.2m telescope.}
{We measure the two-point shear correlation functions to calculate the aperture mass dispersion.
Both statistics are used to perform the data quality control.
Combining the cosmic shear signal with a photometric redshift distribution of a galaxy sub-sample obtained from two square degree of {\it UBVRI}-band observations of the Deep Public Survey we determine constraints for the matter density $\Omega_{\rm m}$, the mass power spectrum normalisation $\sigma_8$ and the dark energy density $\Omega_\Lambda$ in the magnitude interval $R\in [21.5,24.5]$.
In this magnitude interval the effective number density of source galaxies is $n=12.5\,{\rm arcmin}^{-2}$, and their mean redshift is \mbox{$\bar z=0.78$}.
To estimate the posterior likelihood we employ the Monte Carlo Markov Chain method.}
{Using the aperture mass dispersion we obtain for the mass power spectrum normalisation \mbox{$\sigma_8=0.80\pm 0.10$} ($1\,\sigma$ statistical error) at a fixed matter density \mbox{$\Omega_{\rm m}=0.30$} assuming a flat universe with negligible baryon content and marginalising over the Hubble parameter and the uncertainties in the fitted redshift distribution.}
{}

\maketitle


\section{Introduction}
Measuring the weak gravitational lensing effect induced by the tidal gravitational field of the large-scale structure (LSS) of the Universe, also called cosmic shear, is a powerful way to explore statistical properties of the LSS and, thus, cosmological models.
In contrast to other methods cosmic shear directly probes the total matter distribution of the Universe independent of the baryonic content.
Due to the need for high quality data, it has only been in recent years that groups have begun to use this method.
The first cosmic shear measurements were published in the beginning of 2000 \citep{bre00,vme00,kwl00,mvm01,wtk00}.
Although some constraints on cosmological parameters were obtained from these measurements, their statistical errors, due to the small areas of the surveys (\mbox{$\sim 1\,{\rm deg}^2$}), and systematic errors were large.
More recently, published cosmic shear measurements use larger survey areas (\mbox{$\sim 10\,{\rm deg}^2$}) and/or deeper fields to obtain smaller statistical errors and additionally discuss the influence of systematic errors on the cosmic shear signal \citep[e.g.][]{vmr01,vmp02,vmh05,rrg01,rrc04,rrg02,hms03,bmr03,hyg02,btb03,mrb05,jbf03,jjb05,hbb05}.

We are now entering a new phase of wide-field surveys which are wider and/or deeper so that systematic errors may begin to dominate over statistical errors.
Many groups have been working on the problem of reducing systematic errors in cosmic shear measurements and have presented its impact on cosmological parameter estimations.
\citet{hoe04} and \citet{vmh05} discussed the influence of an imperfect PSF-anisotropy correction on the cosmic shear signal.
The impact of the redshift distribution sampling error has recently been studied by \citet{mhh06} and \citet{vwh06}.
A further, more fundamental, source of bias is the correlation between the weak gravitational shear of distant galaxies and the intrinsic shape of foreground galaxies \citep[e.g. ][]{his04,hwh06}.
In addition, the so-called Shear TEsting Programme (STEP), a world-wide collaborative project to improve the accuracy and reliability of all weak lensing measurements has been initiated in mid 2004 \citep[see first results in ][]{hvb06}.

In the present paper we first present the results from a simple test of our pipeline using synthetic images.
Specifically we give details about the changes we have made to it due to the lessons we have learnt so far from STEP and present the influence of the changes on the cosmic shear signal.
To test the pipeline we used $3\,{\rm deg}^2$ of synthetic images obtained by ray tracing through $N$-body simulations \citep{het05}.
A cosmic shear analysis is performed using the E- and B-mode decomposition of the aperture mass ($M_{\rm ap}$-) statistics.
We show that within the uncertainties we recover the cosmic shear signal and that our pipeline does not create artificial B-modes.

We then report on the results of a cosmic shear analysis of the Garching-Bonn Deep Survey (hereafter: GaBoDS) data set, obtained with the Wide Field Imager (WFI) of the MPG/ESO 2.2m telescope on La Silla, Chile.
The GaBoDS data set consists of $18.6\,{\rm deg}^2$ high quality $R$-band observations. 
Most of the data were obtained in the framework of a virtual survey.
That is, we compiled the data set from the ESO archive.
The rest of the data set was obtained by our GO programmes.

We effectively use $15\,{\rm deg}^2$ of the total data set to create catalogues of galaxies for the shear measurements.
Furthermore, the data set includes $2\,{\rm deg}^2$ of deep {\it UBVRI}-band observations from the Deep Public Survey (henceforth: DPS).
This yields photometric redshift information for $8\%$ of all GaBoDS lensing objects in the magnitude interval $R \in [21.5,24.5]$.
The DPS lensing galaxies with photometric redshift information are used to estimate the redshift distribution which in turn is considered for the cosmic shear analysis.

To test the reliability of our measurements we perform four basic tests on the systematics in our data. 
In addition, we calculate the ellipticity cross-correlation between uncorrected stars and corrected galaxies to quantify the residuals of the PSF anisotropy correction.
Furthermore, we utilise the E- and B- mode decomposition of the $M_{\rm ap}$-statistics to check for possible systematics in the data. 

The signal of the aperture mass statistics, the shear two-point correlation functions and their covariance matrix are calculated for the cosmic shear analysis.
In contrast to other cosmic shear analyses, we estimate the mass power spectrum normalisation, $\sigma_8$, using the Monte Carlo Markov Chain (MCMC) technique.

The paper is organised as follows.
In Sect. \ref{sect:cosmicshear} we briefly describe the theory of cosmic shear.
We also discuss the statistics that we use to constrain cosmological parameters as well as a control of possible systematics in the data.
The data and the weak lensing catalogue creation are presented in Sect. \ref{sect:data}.
The analysis of synthetic images is presented in Sect. \ref{sect:simulations}.
In Sect. \ref{sect:Bmodecontrol} we elucidate in detail the different tests on the systematics in our data.
In Sect. \ref{sect:cosmicshearanalysis} we give the results from our cosmic shear analysis and present the cosmological parameters calculated using the MCMC technique.
In addition we discuss several possible sources of systematic errors.
In Sect. \ref{sect:conclusion} we give a summary and outlook.
%
%
\section{Cosmic Shear}
\label{sect:cosmicshear}
In the following we describe the relevant quantities for cosmic shear analyses.
We use the standard lensing notation throughout the paper.
See \citet{bas01} and \citet{sch05} for a more detailed introduction. 
\subsection{Relation between power spectrum and shear estimators}
The power spectrum $P_\kappa$ of the projected density field (convergence $\kappa$) is defined as 
\be
P_\kappa(l)=\frac{9H_0^4\Omega_{\rm m}^2}{4c^4}\int_0^{w_{\rm H}} \dd w \frac{g^2(w)}{a^2(w)}P_{3{\rm D}}\left(\frac{l}{f_K(w)};w\right),
\ee
where $H_0$ is the Hubble constant, $c$ the speed of light, $w$ is the comoving radial coordinate, $a$ is the cosmic scale factor, $P_{3{\rm D}}$ is the three-dimensional mass power spectrum, $l$ is the modulus of the two-dimensional wave vector, and $g$ is the weighting function accounting for the source redshift distribution of galaxies \mbox{$p_z \dd z=p_w \dd w$},
\be
g(w)=\int_w^{w_{\rm H}}\dd w^\prime p_w(w^\prime)\frac{f_{\rm K}(w^\prime-w)}{f_{\rm K}(w^\prime)};
\ee
see for instance \citet{kas98} or \citet{svj98}.
The quantity $f_{\rm K}(w)$ is the comoving angular diameter distance and $w_{\rm H}$ is the distance to the horizon.

In this work we use the shear two-point correlation functions, $\xi_+$, $\xi_-$, and the aperture mass dispersion, $\langle M_{\rm ap}^2\rangle$, to estimate cosmological parameters.
For completeness we also give the shear dispersion, $\langle \gamma^2\rangle$.
All quantities are linearly related to the convergence power spectrum $P_\kappa$ by
\be
\xi_+(\theta_0)=\frac{1}{2\pi}\int_0^\infty \dd l\, l P_\kappa(l)W_{\rm \xi_+}(l\theta_0),
\ee
\be
\xi_-(\theta_0)=\frac{1}{2\pi}\int_0^\infty \dd l\, l P_\kappa(l)W_{\rm \xi_-}(l\theta_0),
\ee
\be
\langle M_{\rm ap}^2\rangle(\theta_0)=\frac{1}{2\pi}\int_0^\infty \dd l\,l P_\kappa(l)W_{\rm ap}(l\theta_0)
\ee
and
\be
\langle \gamma^2\rangle(\theta_0)=\frac{1}{2\pi}\int_0^\infty \dd l\,l P_\kappa(l) W_{\rm TH}(l\theta_0),
\label{eq:tophatshear}
\ee
with 
\be
W_{\rm \xi_+}(\eta)= J_0(\eta), \, W_{\rm \xi_-}(\eta)= J_4(\eta),
\ee
\be
W_{\rm TH}(\eta)=4\frac{J_1^2(\eta)}{\eta^2} \quad {\rm and} \, \quad
W_{\rm ap}(\eta)=576\frac{J_4^2(\eta)}{\eta^4},
\ee
where $J_n$ denote the $n^{\rm th}$-order Bessel function of the first kind.

The aperture mass, $M_{\rm ap}$, was introduced by \citet{sch96_2} and is defined as the spatially filtered projected mass distribution, $\kappa$, inside a circular aperture of angular radius $\theta_0$ at a position $\vek{\zeta}$,
\be 
M_{\rm ap}(\vek{\zeta})\equiv \int\textrm{d}^2\theta\,\kappa(\vek{\theta})\,U(|\vek{\theta}-\vek{\zeta}|),
\ee 
where $U$ is a radially symmetric continuous compensated weight function given in \citet{svj98}.
%
%
The aperture mass $M_{\rm ap}$ can also be expressed in terms of the observable tangential shear $\gamma_{\rm t}$
\be
   M_{\rm ap}(\vek{\zeta})=\int\textrm{d}^2\theta\,\gamma_{\rm t}(\vek{\theta};\vek{\zeta})\,Q(|\vek{\theta}-\vek{\zeta}|),
\label{mapgammat}
\ee
where 
$Q$ is a polynomial filter function given in \citet{svj98}.

The cosmic shear signal can be decomposed into a curl-free (E-mode) and a curl (B-mode) component.
\cite{cnp02} showed that B-modes do not contribute to $M_{\rm ap}$ and E-modes do not contribute to $M_{\perp}$ which is defined as
\be
   M_{\perp}(\vek{\zeta})=\int\textrm{d}^2\theta\,\gamma_{\times}(\vek{\theta};\vek{\zeta})\,Q(|\vek{\theta}-\vek{\zeta}|),
\label{mapcross}
\ee
where
$\gamma_{\times}$ is the cross component of the shear.
As the aperture mass unambiguously separates E- and B-modes, it can therefore be used to reveal possible systematics in the data.

The aperture mass dispersion can directly be obtained from the observed images by placing, for example, a regular grid on the data field.
For each grid point the $M_{\rm ap}$- and $M_{\perp}$-values can be measured and then their dispersion calculated.
However, with this method it is impossible to obtain unbiased estimates for $\langle M_{\rm ap}^2\rangle$ and $\langle M_{\perp}^2\rangle$ because of gaps, holes and borders in the data fields.
We therefore calculate the dispersions directly in terms of the shear two-point correlation functions, which are directly related to the observable quantities $\gamma_{\rm t}$ and $\gamma_\times$:
\be
\xi_\pm(\theta) = \langle\gamma_{\rm t}\gamma_{\rm t}\rangle(\theta) \pm \langle\gamma_\times \gamma_\times\rangle (\theta),
\ee
\be
\xi_\times(\theta) = \langle \gamma_{\rm t}\gamma_{\times}\rangle(\theta).
\ee
Due to parity invariance the correlation function $\xi_\times$ is expected to vanish.

The aperture mass dispersion is related to the shear two-point correlation functions, $\xi_\pm$, by
\be
\langle M_{\rm ap}^2\rangle\!(\theta_0\!)\! = \! {1\over 2}\int \! {\dd \theta\,\theta \over \theta_0^2}\!
\left[\xi_+(\theta)\,T_+\!\!\left(\! \theta \over \theta_0 \! \right)\!+\!\xi_-(\theta)\,T_-\!\!\left(\! \theta \over \theta_0 \! \right)\right]
\label{eq:mapdis}
\ee
\be
\langle M_{\perp}^2\rangle\!(\theta_0\!)\! = \! {1 \over 2}\int \! {\dd \theta\,\theta \over \theta_0^2}\!
\left[\xi_+(\theta)\,T_+\!\!\left(\! \theta \over \theta_0 \! \right)\!-\!\xi_-(\theta)\,T_-\!\!\left(\! \theta \over\theta_0 \! \right)\right]\!.
\label{eq:Bmapdis}
\ee
Another quantity defined by
\be
\langle M_\times^2 \rangle (\theta_0)=\int {\dd \theta\,\theta \over \theta_0^2}\,\xi_\times(\theta)\,T_-\!\!\left(\! \theta \over\theta_0 \! \right)
\label{eq:map2cross}
\ee
can also be used as an indicator for systematics as it is unaffected by cosmic shear.
The functions $T_\pm$ can be expressed analytically \citep{svm02} and vanish by definition for $\theta>2\,\theta_0$.
Thus $\langle M_{\rm ap}^2\rangle$, $\langle M_{\perp}^2\rangle$ and $\langle M_\times^2 \rangle$ can be obtained directly from the correlation functions $\xi_\pm$ in a finite interval.
In Eq. (\ref{eq:map2cross}) we choose $T_-$ because it has no change of sign in contrast to $T_+$.

In this work we utilise the aperture mass dispersion to separate E- and B-modes \citep[as suggested in ][]{svm02}.
Throughout this paper we use $\langle M_{\rm ap}^2\rangle$ and E interchangeably as well as $\langle M_{\perp}^2\rangle$ and B.
\subsection{Shear estimators in practice}
The estimators of the two-point shear-shear correlation functions in practice are
\be
\hat\xi_\pm(\theta)=\frac{\sum_{i,j}^N w_i w_j (\epsilon_{{\rm t},i}\epsilon_{{\rm t},j} \pm \epsilon_{\times,i}\epsilon_{\times,j})\Delta_{ij}(\theta)}{\sum_{i,j}^N w_i w_j \Delta_{ij}(\theta)},
\ee
with
\begin{displaymath}
\Delta_{ij}(\theta)=\left\{
\begin{array}{ll}
1,\, & \theta-\Delta\theta/2 \leq |\theta_i-\theta_j|<\theta+\Delta\theta/2 \\
0,\, & {\rm otherwise},
\end{array}
\right.
\end{displaymath}
where $\epsilon_{\rm t}$ and $\epsilon_{\times}$ are the tangential and cross components of the corrected galaxy ellipticities and $w_i$ is the weighting factor (details of the correction and weighting scheme can be found in Sect. \ref{sec:psfcorrection} and \ref{sec:catalogue}).
All pairs of background galaxies within a distance of $\theta-\Delta\theta/2$ and $\theta+\Delta\theta/2$ are considered,
where $\Delta\theta$ is the angular width of the bin.
The practical estimators of the aperture mass dispersion are
\begin{eqnarray}
\hat M(\theta_0)\!&=&\!{\Delta\theta \over 2\theta_0^2}\sum_{n=1}^{2m}\theta_n\!\left[ \hat\xi_+(\theta_n)\,T_{\!+}\!\left(\theta_n\over\theta_0\!\right)\!+\!\hat \xi_-(\theta_n)\,T_{\!-}\!\left(\theta_n\over\theta_0\!\right)\right]\!,\nonumber\\
\hat M_\perp(\theta_0)\!&=&\!{\Delta\theta \over 2\theta_0^2}\sum_{n=1}^{2m}\theta_n\!\left[ \hat\xi_{\!+}(\theta_n)\,T_+\!\left(\theta_n\over\theta_0\!\right)\!-\!\hat \xi_{\!-}(\theta_n)\,T_-\!\left(\theta_n\over\theta_0\!\right)\right]\!,\nonumber\\
\hat M_\times(\theta_0) & = & {\Delta\theta \over \theta_0^2}\sum_{n=1}^{2m}\theta_n\, \hat\xi_\times(\theta_n)\,T_-\left(\theta_n\over\theta_0\!\right),
\end{eqnarray}
with the centre of the bins $\theta_n=(n-1/2)\Delta\theta$ and the aperture radius $\theta_0$.

%
%
\section{Data and galaxy catalogues}
\label{sect:data}
\subsection{The GaBoDS data set}
\label{sec:dataset}
The Garching-Bonn Deep Survey (GaBoDS) data set was obtained with the Wide Field Imager (WFI) of the MPG/ESO 2.2m telescope on La Silla, Chile.
The camera consists of eight $2\,{\rm k}\times 4\,{\rm k}$ CCDs with a pixel size of $15\,\mu {\rm m}$, corresponding to a pixel scale of $0\myarcsec 238$ in the sky.
The field-of-view is $34^\prime \times 33^\prime$.
However, due to the applied dither pattern the effective field-of-view can be as large as $40^\prime \times 40^\prime$.

The total data set consists of $18.6\,{\rm deg}^2$ of high quality $R$-band observations.
About $80\%$ of the data were obtained by a virtual survey (ASTROVIRTEL project\footnote{ASTROVIRTEL cycle 2: Erben et al.,Gravitational lensing studies in randomly distributed, high galactic latitude fields}$^,$\footnote{http://www.stecf.org/astrovirtel}), meaning that the data set was collected from the ESO archive.
For data mining the archive, the search engine {\tt querator}\footnote{http://archive.eso.org/querator} was developed by \citet{pie01} to address the needs of astronomers looking for multicolour imaging data. 
Fields were selected by four criteria: (1) the exposure time was at least $3.5\,{\rm ks}$, (2) the seeing was better than $1^{\prime \prime}$, (3) the field was not affected by bright foreground objects and (4) neither the field nor adjacent fields contained previously known very massive structures.
For further details of the data mining of the ESO archive we refer to \citet{ses03}.
The rest of the data set was obtained by our GO programmes (table \ref{SumTab}).

The data set is heterogeneous with respect to the Vega-limiting magnitude in the $R$-band (ranging between $25.0\,{\rm mag}$ and $26.5\,{\rm mag}$; $5\,\sigma$ sky level measured in a circular aperture of $2^{\prime \prime}$ radius), seeing (ranging between $0\myarcsec 75$ and $1\myarcsec 20$), observing strategy (dithering/no dithering) and co-addition process (different programmes were used for co-addition: {\tt EISdrizzle} and {\tt swarp}\footnote{\citet{esd05} showed that there is no significant difference in the size of the PSF or the PSF anisotropy pattern using {\tt EISdrizzle} or {\tt swarp}.}, see table \ref{tab:datasource}).
The positions of the fields are randomly distributed on the southern hemisphere at high galactic latitudes, however some of the fields form small patches in the sky.
The survey is separated into six data sets mainly depending on the data source, see table \ref{tab:datasource} and \ref{SumTab}.
Note that the ESO Distant Cluster Survey \citep[EDisCS,][]{csa06,wcs06} data set contains distant galaxy cluster candidates ($z=0.5...0.8$) that have been identified using the red cluster sequence.
Since we perform our lensing analysis with those objects lying in the magnitude interval $R \in [21.5,24.5]$, we are presumably not sensitive to those high redshift clusters.
We assume in the following that these fields do not bias our cosmic shear survey significantly towards high-density regions.

Each of the WFI images observed with dither pattern ($\approx 70\,\%$ of all fields) cover an area of about $0.35\,{\rm deg}^2$.
However, for the cosmic shear analysis the edges are trimmed off (with a low effect on the total number density of background sources) resulting in an effective area of $0.25\,{\rm deg}^2$ per field.

Furthermore, the data set includes $2\,{\rm deg}^2$ of deep {\it UBVRI}-band observations from the DPS \citep{hed05}, which yields photometric redshift information for $8\%$ of the objects considered for the cosmic shear analysis.
\subsection{Data reduction}
The data reduction has been performed with a nearly fully automated, stand-alone pipeline which has been developed by our group to reduce optical and near-infrared images, especially those taken with multi-chip cameras.
A detailed description of the pipeline can be found in \citet{esd05}.
As weak gravitational lensing was the main science driver, the pipeline algorithms are optimised to produce deep co-added mosaics from individual exposures obtained from empty field observations.
Special care has been taken to achieve an accurate astrometry to reduce possible artificial PSF patterns in the final co-added images.
Since keeping the PSF anisotropies small is mandatory for cosmic shear measurements, individual exposures are rejected from the co-addition process if one or more CCDs exhibit an anisotropy larger than $6\,\%$ in either of the ellipticity components.
More information about the exact data reduction process can be found in \citet{sch04}.
Note that for the image co-addition of the DPS data set (code name in our work: DEEP, table \ref{tab:datasource}), the programme {\tt swarp} was used in contrast to the rest of the GaBoDS data set that was co-added with {\tt EISdrizzle}.
\begin{table}
\caption{Data code, depth of the fields (time $t$ of all co-added exposures), total field area, effective number density $n_{\rm eff}$ and data source (see table \ref{SumTab} for details of individual fields). 
Key:
$^*$: co-addition with {\tt swarp},
$^\dagger$: single exposures are not dithered,
$^\ddagger$: fields are rejected for the final cosmic shear analysis.}
\label{tab:datasource}
\begin{center}
    \begin{tabular}{l|c|r|r|l}
      \hline \hline
      code             & depth t        & {\centering area}           &$n_{\rm eff}$&Data source \\
                       & [ks]           & {\centering [${\rm deg}^2$]}&$[{^\prime}^{-2}]$& \\
      \hline
      COMBO            & $10\!<\!t\!<\!56$   & 1.25                   &16.5 &COMBO-17 \\
      \hline
      OWN              & $7.0 \!<\!t\!<\! 22$  & 3.25                 &12.0 & own obs., \\
      &                                 &                             &     & MPE IR group,\\
      &                                 &                             &     &ESO archive \\
      \hline
      ${\rm DEEP}^*$             & $3.9 \!<\!t\!<\! 9.3$ & 3.25       &11.5 & EIS + \\
      &                                 &                             &     &own obs. \\
      \hline
      ${\rm B8}^\dagger$& $4.2 \!<\!t\!<\! 7.5$ & 2.5                  &11.5 & ASTROVIRTEL \\
      \hline
      ${\rm C0}^{\dagger,\,\ddagger}$& $4.0 \!<\!t\!<\!4.8$ & 2.0       &10.0 & ASTROVIRTEL \\
      \hline
      CL               & $t=3.6$        & 3.0                         &12.0  & EDisCS \\
      \hline
    \end{tabular}
\end{center}
\end{table}
\subsection{PSF correction}
\label{sec:psfcorrection}
The shape of galaxies is influenced by the anisotropic PSF. 
In order to obtain a correct estimate of the shear $\gamma$ from the observed ellipticity of galaxies $\chi^{\rm obs}$, \cite{ksb95} developed the so-called KSB algorithm.
The algorithm relates the observed ellipticities $\chi^{\rm obs}$ to the sheared source-ellipticities and provides an unbiased estimate of the shear.
The correction is calculated on the Gaussian kernel weighted second brightness moments $Q_{ij}$ of a galaxy with surface brightness $I(\vek{\theta})$,
\begin{equation}
Q_{ij}=\int {\rm d}^2\theta \,(\theta_i-\bar\theta_i)(\theta_j-\bar\theta_j)\,I(\vek{\theta})\,W\left(\left|\vek{\theta}-\vek{\bar\theta}\right|^2\right),
\label{eq:quadro}
\end{equation}
with 
\be
W\left(|\vek{\theta}-\vek{\bar\theta}|^2\right)=\frac{1}{2\pi \sigma^2}{\rm exp}\left(-\frac{|\vek{\theta}-\vek{\bar\theta}|^2}{2\sigma^2}\right),
\label{eq:gewicht}
\ee
where the scale length, $\sigma$, is defined as the half-light radius, $r_{\rm h}$, of the object considered, and the centre is 
\be
\vek{\bar\theta}\equiv \frac{\int\dd^2\theta\, W(\theta)\,\vek{\theta}\,I(\vek{\theta})}{\int\dd^2\theta\, W(\theta)\,I(\vek{\theta})}.
\label{eq:centre}
\ee
The ellipticity is defined as
\begin{equation} 
\chi \equiv \frac{Q_{11}-Q_{22}+2{\rm i}Q_{12}}{Q_{11}+Q_{22}}.
\label{elli}
\end{equation}

Assuming that the intrinsic orientation of galaxies is random, the relation between $\gamma$ and $\chi^{\rm obs}$ in the weak lensing regime reads
\be
\gamma= (P^{\rm g})^{-1}(\chi^{\rm obs}-P^{\rm sm}q^*),
\label{eq:pgcorr}
\ee
where $P^{\rm g}$ is the pre-seeing shear polarisability which depends on the smear and shear polarisability tensors $P^{\rm sm}$ and $P^{\rm sh}$ of the galaxy, and the stellar smear and shear polarisability tensors $P^{{\rm sm}*}$ and $P^{{\rm sh}*}$ in the following way
\be
P^{\rm g}=P^{\rm sh}-P^{\rm sm}(P^{{\rm sh}*})^{-1}P^{{\rm sm}*}.
\ee
The tensor $P^{\rm g}$ relates the measured anisotropy-corrected galaxy ellipticity to its true value.
All quantities are calculated by means of the observable surface brightness, $I(\vek{\theta})$.
The quantity $q^*$ is the anisotropic part of the PSF and is calculated from the raw stellar ellipticity $\chi^*$, $q^*=(P^{\rm sm *})^{-1}\chi^*$.
\subsection{Catalogue creation}
\label{sec:catalogue}
\paragraph{Raw catalogue. }
We utilise the programme {\tt SExtractor} \citep{bea96} to create a raw catalogue of objects (source galaxy candidates).
The final co-added image is first smoothed with a Gaussian kernel of $2.5\,{\rm pixel}$ FWHM (FILTER\_NAME: gauss\_2.5\_5x5.conv).
Then objects are extracted which consist of at least \mbox{$N=5$} contiguous pixels ({\tt SExtractor} parameter `DETECT\_MINAREA') with a flux greater than \mbox{$k=1.5\,\sigma$} above the sky level noise ({\tt SExtractor} parameter `DETECT\_THRESH').
With these rather conservative parameters we reduce the number of noise detections and increase the reliability of shape measurements. 
For the object detection the co-added WEIGHT image is taken into account as an additional {\tt SExtractor} argument.
It provides a full characterisation of the relative noise properties for each science image pixel and therefore lowers the probability to detect pure noise objects, see \citet{esd05}.
In addition, we set BACK\_TYPE=MANUAL and BACK\_VALUE=0 (our co-added images are sky subtracted).
In this way {\tt SExtractor} does not model halos around bright objects as sky background.
It turns out that this effectively reduces spurious detections of objects in the proximity of bright stars or within their halos. 
\paragraph{Calculation of quadrupole moments, $Q_{ij}$ and centroids, $\vek{\bar\theta}$. }
The raw catalogue is transfered to the programme {\tt analyseldac}, an adjusted version of N. Kaisers {\tt imcat}, which measures the quadrupole moments of each object detected with {\tt SExtractor}.
Due to pixelisation, the continuous integrals of Eqs. (\ref{eq:quadro}) and (\ref{eq:centre}) are now transformed into discrete sums,
\be
Q_{ij}=\!\!\sum_{\theta_i,\theta_j=-\theta_{\rm max}}^{\theta_{\rm max}}\!\!\Delta\theta^2\,
       (\theta_i-\bar\theta_i)(\theta_j-\bar\theta_j)\,
       I(\vek{\theta})\,
       W\left(\left|\vek{\theta}-\vek{\bar\theta}\right|^2\right)
\label{eq:discreteQ}
\ee
and
\be
\vek{\bar\theta}= \frac{\sum_{\theta_i,\theta_j=-\theta_{\rm max}}^{\theta_{\rm max}} \Delta\theta^2\, 
                  W(|\vek{\theta}-\vek{\bar\theta}|)\, \vek{\theta}\, I(\vek{\theta})}
                  {\sum_{\theta_i,\theta_j=-\theta_{\rm max}}^{\theta_{\rm max}} \Delta\theta^2\,
		  W(|\vek{\theta}-\vek{\bar\theta}|)\, I(\vek{\theta})},
\label{eq:discretecentre}
\ee
where $\theta$ is measured, in pixel units, from the source centroid $\vek{\bar\theta}$ and $\Delta \theta=0.25\,{\rm pixel}$ (each pixel is divided into four sub-pixels).
The object centre $\vek{\bar\theta}$ is determined by iteratively solving Eq. (\ref{eq:discretecentre}), where the starting point is the {\tt SExtractor} centroid. 
The scale in the Gaussian weighting function, $W$, is the half-light radius, $r_{\rm h}$, which is equal to the {\tt SExtractor} parameter FLUX\_RADIUS.
The quadrupole moments, $Q_{ij}$, are calculated from all pixels for which \mbox{$|\vek{\theta}-\vek{\bar\theta}|<\theta_{\rm max}=3\,r_{\rm h}$}. 
For non integer pixel values of the position $\vek{\theta}$, the surface brightness $I(\theta_i,\theta_j)$, known at pixel positions, is estimated from a linear interpolation over the four nearest pixels to $(\theta_i,\theta_j)$. 
\paragraph{PSF-anisotropy correction.  }
In order to apply Eq. (\ref{eq:pgcorr}) the anisotropy term, $q^*$, has to be known.
For the PSF-anisotropy correction we therefore utilise stars which are point-like and unaffected by lensing.
All stars have the same half-light radius and therefore show up as a vertical branch in a magnitude against half-light radius plot.
Stars which have a brightness of \mbox{$R=0.5$} magnitudes lower than saturated stars and which are well above the crowded faint magnitude region which contains a mixture of faint stars, galaxies and noise detections are selected.
Using this sample of bright, unsaturated stars, we measure $q^*$ with a Gaussian filter scale matched to the size of the galaxy image to be corrected.
\citet{hfk98} proposed this matched PSF-anisotropy correction scheme for space-based data.
It turned out in STEP 2 (Massey et al., in preparation) that this also improves the results for ground-based data.

In the case of the WFI@2.2m instrument the PSF anisotropy of the co-added images is rather small and varies smoothly over the total field-of-view.
Discontinuities in the PSF anisotropy are largely absent across chip borders \citep[demonstrated in][]{ses03,sch04,esd05}.
Therefore we perform a second- or third-order two-dimensional polynomial fit to $q^*$ with $3.5\,\sigma$-clipping as a function of position over the entire field-of-view. 
With this fit it is possible to estimate the anisotropy kernel, $q^*$, at the position of the galaxies.
A basic analysis of PSF anisotropy properties of the GaBoDS data set is given in Schirmer et al. (in preparation).
Note that this anisotropy correction method is not perfect as the PSF is only measured at the position of stars and a polynomial {\it interpolation} is used to estimate the PSF anisotropy at the position of galaxies.
However, our fields comprises several hundreds to some thousands of stars used for the anisotropy correction, so that the PSF is sampled very well.

For future cosmic shear analyses aiming at high precision constraints for cosmological parameters the PSF anisotropy correction method must be improved.
Recently \citet{jaj04} have introduced a promising correction algorithm, the so-called principal component analysis.
With this method the dominant PSF anisotropy patterns are identified from a large number of exposures and are used to model the PSF.
In this way, the PSF anisotropy correction does no longer depend on the number density of stars per field, but by the stacked number density of stars across all fields of the survey. 
\paragraph{Calculation of $P^{\rm g}$ and $\gamma$.  }
The $P^{\rm g}$ tensor is an extremely noisy quantity.
The diagonal elements of all required tensors to calculate $P^{\rm g}$ are dominant by at least a factor of 10 compared to the off-diagonal elements, and they are approximately equal, so that we can estimate $P^{\rm g}$ by $P^{\rm g}_{\rm s}\mathbbm{1}$, with 
\be
P^{\rm g}_{\rm s}=\frac{1}{2}\,\left\{{\rm tr}(P^{\rm sh})-\frac{{\rm tr}(P^{\rm sh *})}{{\rm tr}(P^{\rm sm *})}\,{\rm tr}(P^{\rm sm})\right\},
\ee
see \citet{ewb01}.
The stellar smear and shear polarisability tensors $P^{{\rm sm} *}$ and $P^{{\rm sh} *}$ are calculated for different smoothing scales $r^*_{\rm h}$ and a fifth-order polynomial fit as a function of position is performed to ${\rm tr}(P^{{\rm sm} *})/{\rm tr}(P^{{\rm sh} *})$ since it varies over the total field-of-view. 
As $P^{\rm g}$ depends on these stellar quantities we calculate $P^{\rm g}$ depending on the galaxy size, $r_{\rm h}$.
For $P^{\rm g}$ we use the raw, unsmoothed values because \citet{ewb01} pointed out that fitting or calculating means does not improve the shear estimates.

Our final shear estimate is given by \mbox{$\gamma_{\rm corr}=\gamma/f_{\rm cal}$}, where $\gamma$ is calculated according to Eq. (\ref{eq:pgcorr}) and \mbox{$f_{\rm cal}=0.88$} is a calibration factor obtained from {\tt skymaker} and shapelet simulations \citep[see STEP 1, 2: ][ Massey et al. in preparation]{hvb06}.  
%
%
%
\begin{figure}
\centering
\includegraphics[scale=0.35,angle=-90]{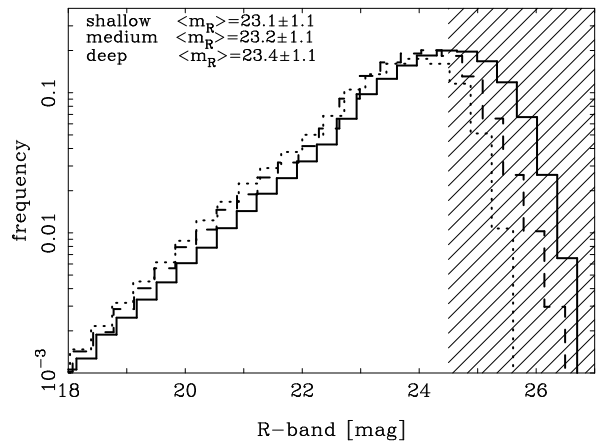}
\caption{
Galaxy counts in $R$-band magnitude of the GaBoDS for three different exposure time intervals.
Solid line: deep ($10\,{\rm ks}<t<56\,{\rm ks}$),
dashed line: medium ($7\,{\rm ks}<t<10\,{\rm ks}$),
dotted line: shallow ($3.6\,{\rm ks}<t<7\,{\rm ks}$).
All distribution functions are normalised by area between $18 < R < 24.5$.
Galaxies with $R>24.5$ (shaded area) are not considered in this work.
}
\label{fig:completeness}
\end{figure}
\paragraph{Catalogue filtering.  }
All objects for which problems concerning the determination of shape or centroid position occur are rejected
(e.g., objects near the border, with negative total flux, with negative \mbox{$Q_{11}+Q_{22}$}, with negative semi-major and/or semi-minor axis, or when the iterative centroid determination is not stable after a certain number of steps).
In addition, we only use those objects with a half-light radius which is larger than that measured for stars and a modulus of the ellipticity (after PSF correction) of less than 1.0.

\begin{figure}
\centering
\includegraphics[scale=0.42]{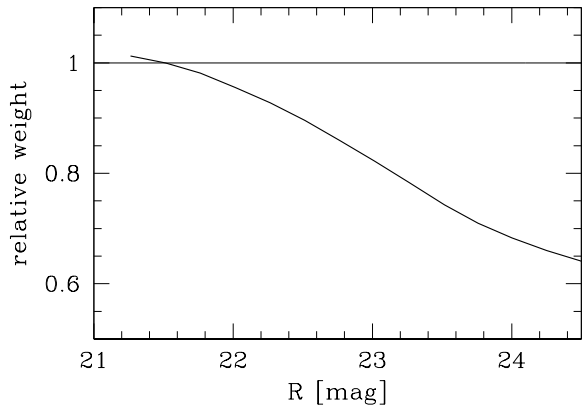}
\caption{
Average weighting factor calculated from all galaxies as a function of magnitude.
The weighting factors are calculated with Eq. (\ref{eq:weight}) such that galaxies with a magnitude of $R=21.5$ have a weight of 1.
}
\label{fig:galaxweight}
\end{figure}

To obtain a more homogeneous data set, with respect to the completeness of the object catalogue of individual fields, the raw background galaxy catalogues are compiled from objects in the {\tt SExtractor} isophotal magnitude interval $R \in [21.5,24.5]$.
In this magnitude range we are confident to obtain reasonable photometric redshift estimates for most of the lensing galaxies in the individual DPS fields so that we have a reliable redshift distribution for all lensed galaxies. 
In Fig. \ref{fig:completeness} the galaxy number counts are displayed.
The distributions are approximately equal below $R<24.5$, independent of the exposure time so
we perform a cut at a magnitude where completeness stops for all fields.
Hence we would not improve our final cosmic shear analysis significantly by going deeper given that faint galaxies are strongly downweighted anyway (Fig. \ref{fig:galaxweight}). 
%
%
%
\paragraph{Weighting.  }
For each galaxy the next 12 neighbours are identified in the magnitude-half-light radius plane and the variance, $\sigma_e^2$, of the ellipticity distribution of the sub-sample is calculated \citep[for more details about this technique see][]{ewb01}.
The variance, $\sigma_e^2$, gives an indication of the noise level of this galaxy.
From $\sigma_e^2$ we then determine the weighting factor, $w_i$, as,  
\be
w_i=\frac{1}{\sigma_e^2+\bar \sigma^2}
\label{eq:weight}
\ee
where $\bar \sigma^2=0.16$ is the variance of the unlensed galaxies (estimated from the total PSF corrected galaxy sample).
In Fig. \ref{fig:galaxweight} the mean weighting factor is displayed as a function of magnitude.
In the magnitude interval considered, we find that the average relative weight measured by \citet{hmv05} is comparable to ours, although they use a completely different weighting scheme.
%
%
\section{Cosmic shear analysis of synthetic images}
\label{sect:simulations}
In this section we present the results from qualitative tests of our pipeline.
For simplicity we divide our pipeline into two parts: (1) the part concerning the PSF correction and shear estimation and (2) the part which measures the two-point correlation function and calculates the aperture mass dispersion. 
In addition, we investigate the impact of the changes we have made to part (1) of our pipeline (see table \ref{tab:changes}) on the cosmic shear signal due to the lessons we have learnt so far from STEP. 

We performed a cosmic shear analysis of twelve synthetic WFI sized images created from ray tracing through $\Lambda$CDM $N$-body simulations (\mbox{$\Omega_\Lambda=0.7,\, \Omega_0=0.3,\, \sigma_8=0.9,\, h=0.7$}) using the E- and B-mode decomposition of the $M_{\rm ap}$-statistics.
These simulations were kindly made available by Takashi Hamana [details see \cite{hty04}].
We created twelve initial catalogues of randomly distributed galaxies using the programme {\tt stuff}\footnote{Available at:\\ \textup{http://terapix.iap.fr/cplt/oldSite/soft/stuff/}} (E. Bertin).
The galaxies are assumed to be at a fixed redshift $z=1$, and are sheared according to the shear map of the ray-tracing simulations.
In the following these galaxy catalogues are called \textit{\textbf{input catalogues}}. 

The input catalogues are used to create synthetic images using the programme {\tt skymaker}\footnote{Available at:\\ \textup{http://terapix.iap.fr/cplt/oldSite/soft/skymaker/}} by E. Bertin.
A short description is given in \cite{ewb01}.
Twelve $30^\prime\times 30^\prime$ images resulting in a $3\,{\rm deg}^2$ survey are obtained for the twelve catalogues.  
These images are treated in exactly the same way as real data (like object detection, PSF correction, same cuts, weighting), see Sect. \ref{sec:catalogue}.
The obtained galaxy catalogues are called \textit{\textbf{output catalogues}}.

To exclude the effect of false detections we only take into account those objects which are present in both, input and output catalogues. 
To obtain a similar number density in comparison to the real data we perform a similar magnitude cut: $mag \in [21.5,24.5]$. 
The mean ellipticity dispersion of the galaxies of the input and output catalogues is $\sigma_\epsilon=0.32$ and the mean galaxy number density is $n=15\,{\rm arcmin}^{-2}$.
For a more detailed description of the image creation, galaxy morphology and magnitude distribution see \cite{het05} and \cite{ewb01}.
\subsection{Noise-free case}
\label{sec:noise-free}
\begin{figure}
\centering
\includegraphics[scale=0.4]{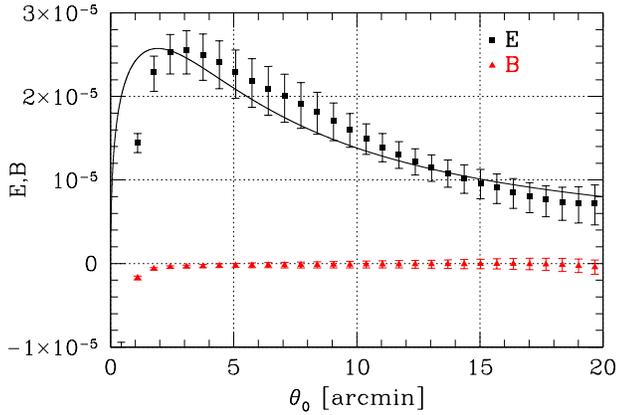}
\caption{
Noise-free case: cosmic shear analysis of the input catalogue.
The number density of background sources is \mbox{$n=15\,{\rm arcmin}^{-2}$}, and the sources are intrinsically round.
The signal in each bin is the average of the 12 fields, where the statistical weight for each field is unity; the error bars are obtained by bootstrapping, see Sect. \ref{para:boot2} and are highly correlated.
Displayed is the aperture mass dispersion (E- and B-modes).
The line is a $\Lambda$CDM prediction assuming $\Omega_{\rm m}=0.3$, $\Omega_\Lambda=0.7$, $\Gamma=0.21$ and $\sigma_8=0.9$ and $z_{\rm source}=1$ (Peacock \& Dodds model of the power spectrum).
}
\label{fig:bulge_theory}
\end{figure}

To neglect the intrinsic shape of galaxies we perform a cosmic shear analysis of the input catalogue for intrinsically round objects using the aperture mass statistics.
The results are given in Fig. \ref{fig:bulge_theory}.
The displayed $\Lambda$CDM prediction is based on the \citet{ped96} model of the non-linear power spectrum.
The cosmic shear signal is in excellent agreement with the theoretical prediction.
Hence we conclude that the pipeline measures the two-point correlation function and calculates the aperture mass dispersion exactly, and does not create artificial B-modes.

For angular scales smaller than \mbox{$\theta_0 < 2\,{\rm arcmin}$} the E- and B-mode signals are significantly smaller than expected.
This effect is explained and quantified in \citet{kse06}.
The measured two-point correlation functions, $\xi_\pm$, are set to zero for small angular scales (here we choose the same value as for the observations: \mbox{$\theta_0<6\,{\rm arcsec}$}) because close galaxy pairs are rejected.
This lack of shear correlation measurements on small scales results in a mixing of E- and B-modes.

Note that for angular scales larger than two arcminutes the marginal difference between the theoretical prediction and simulations can easily arise from cosmic variance due to the fact that we only use $3\,{\rm deg}^2$ from the numerical simulations.
%
%
%
\subsection{Realistic case}
Using the ``old'' pipeline the shear measurements of simple {\tt skymaker} simulations yielded, on average, an underestimation of the shear by \mbox{$\sim 15\,\%$} \citep[STEP 1, see][]{hvb06}.
With our (modified) pipeline we are currently able to recover the shear with less than $\approx 3\,\%$ calibration bias (depending on the PSF model).
This has been checked with a second set of more realistic shapelet-simulations (STEP 2: Massey et al. in preparation).
A short comparison between the modified and ``old'' pipeline is given in table \ref{tab:changes}.

\begin{figure}
\centering
\includegraphics[scale=0.4]{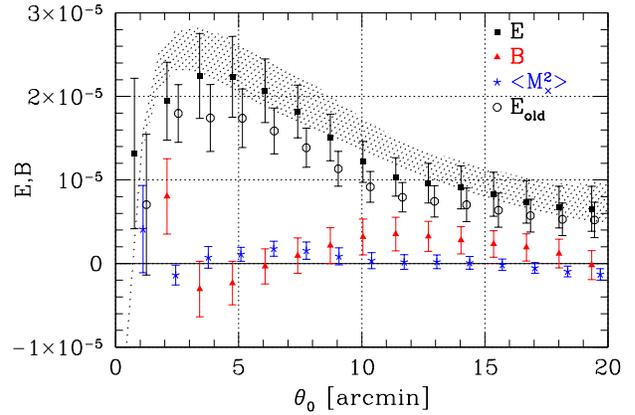}
\caption{
Realistic case: cosmic shear analysis of the output catalogue. 
The same objects are used as in Fig. \ref{fig:bulge_theory}. 
Shaded area: $1\,\sigma$ range of the signal in Fig. \ref{fig:bulge_theory}.
Displayed are the aperture mass dispersion (E- and B-modes) and $\langle M^2_\times\rangle$. 
In addition, the ${\rm E}_{\rm old}$-mode signal obtained via the ``old'' pipeline prior STEP 1 is displayed.
}
\label{fig:bulge_real}
\end{figure}

The same analysis as in Sect. \ref{sec:noise-free} is performed for the output catalogue.
We present the results of the analysis done with the ``old'' and modified pipeline, prior and after the changes we have made to our pipeline due to STEP 1 and 2.
Both results are compared with the expected signal of the noise-free case, see Fig. \ref{fig:bulge_real}.
Clearly visible is the improvement of the modified pipeline compared to the ``old'' pipeline.
However, we still slightly underestimate the cosmic shear signal with the modified pipeline by a few percent. 
In addition, our pipeline does not create significant artificial B-modes (note that the error bars in Fig. \ref{fig:bulge_real} are strongly correlated and the B-mode signal varies around zero).

We conclude that with our updated cosmic shear pipeline including shear estimation and calculation of various cosmic shear statistics we can perform accurate measurements of the cosmic shear signal within the currently attainable accuracy. 
\begin{table}
\caption{Comparison between the modified and ``old'' pipeline.
Both pipelines are based on the KSB algorithm outlined in the text.
The changes made to the pipeline are based on results given in \citet{hvb06} and Massey et al. (in preparation).
Key:
$^\dagger$: this change was made after STEP 2.
}
\label{tab:changes}
\begin{center}
    \begin{tabular}{l|l}
      \hline \hline
       modified pipeline                           & ``old'' pipeline               \\
       \hline
       anisotropy measurement$^\dagger$:                      & \\ 
       $q^*$  matched with a Gaussian              & fixed Gaussian filter      \\
       filter scaled to the size of the            & scale                      \\
       galaxy image to be corrected                &                             \\
       \hline
        cuts:                                       &  \\  
       $|\gamma|<1.0$                              & $|\gamma|<0.8$               \\
       ${\rm tr}[P^{\rm g}]>0.05$                   & ${\rm tr}[P^{\rm g}]>0$\\
       \hline
       calibration factor $\gamma_{\rm corr}=\gamma/0.88$& no calibration factor        \\
       \hline
    \end{tabular}
\end{center}
\end{table}
%
%
%

%
%
\section{Dealing with systematics in the GaBoDS data}
\label{sect:Bmodecontrol}
In this section we describe various tests to discover systematic errors in our data set.
First we present four simple and quick quality checks of the data.
Second, we present a powerful standard test for systematics, the cross-correlation between uncorrected stars and anisotropy corrected galaxies \citep{bmr03}.
All tests described basically provide checks on the quality of the PSF anisotropy correction.
However, to test the smearing correction (i.e. the correction for the PSF size) we have to rely on simulations.
For that purpose the aforementioned STEP project has been initiated.
\subsection{Four basic quality tests}
\label{quality}
\begin{figure}
\centering
\includegraphics[scale=0.43]{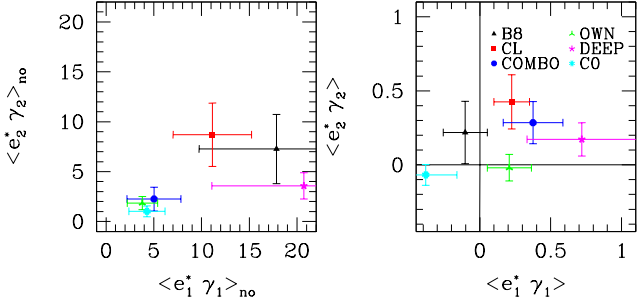}
\caption{
Test of the anisotropy correction in the magnitude range $R \in [21.5,24.5]$ using Eq. (\ref{eq:crossmultiply}).
{\bf Left panel: } before anisotropy correction; {\bf right panel: } after anisotropy correction.
Note the drastic change of scale.}
\label{fig:anitest}
\end{figure}

\begin{figure}
\centering
\includegraphics[scale=0.45,clip]{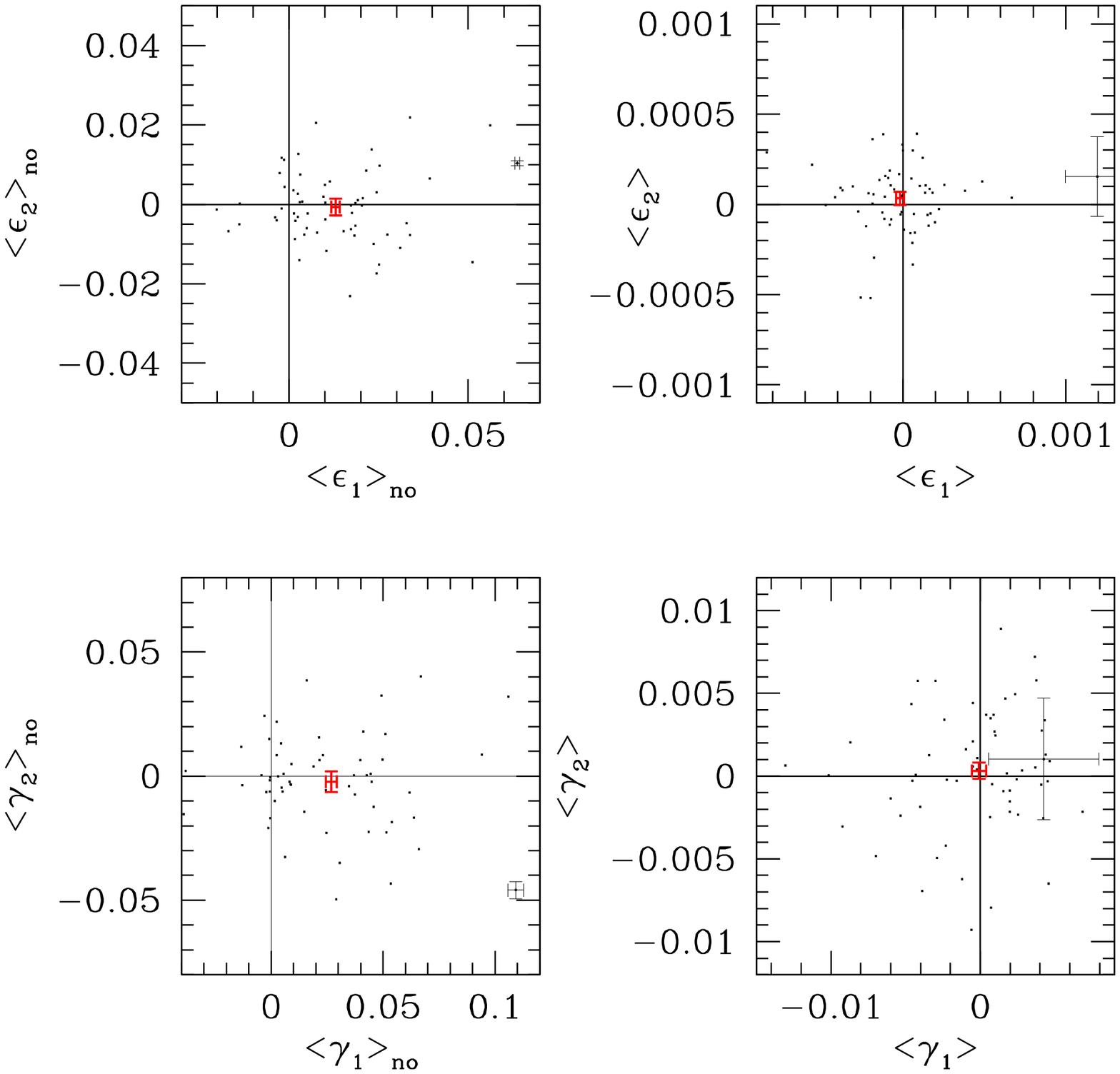}
\caption{
Average shear of all galaxies in the considered magnitude bin $R \in [21.5,24.5]$ calculated for all GaBoDS fields before ({\bf left panel}) and after anisotropy correction ({\bf right panel}). 
Because of clarity we only plot one error bar for a single field (thin lines).
The average of all fields is plotted with a solid error bar.
Note the drastic change of scale.
}
\label{fig:ani_shear}
\end{figure}

\begin{figure}
\centering
\includegraphics[scale=0.32]{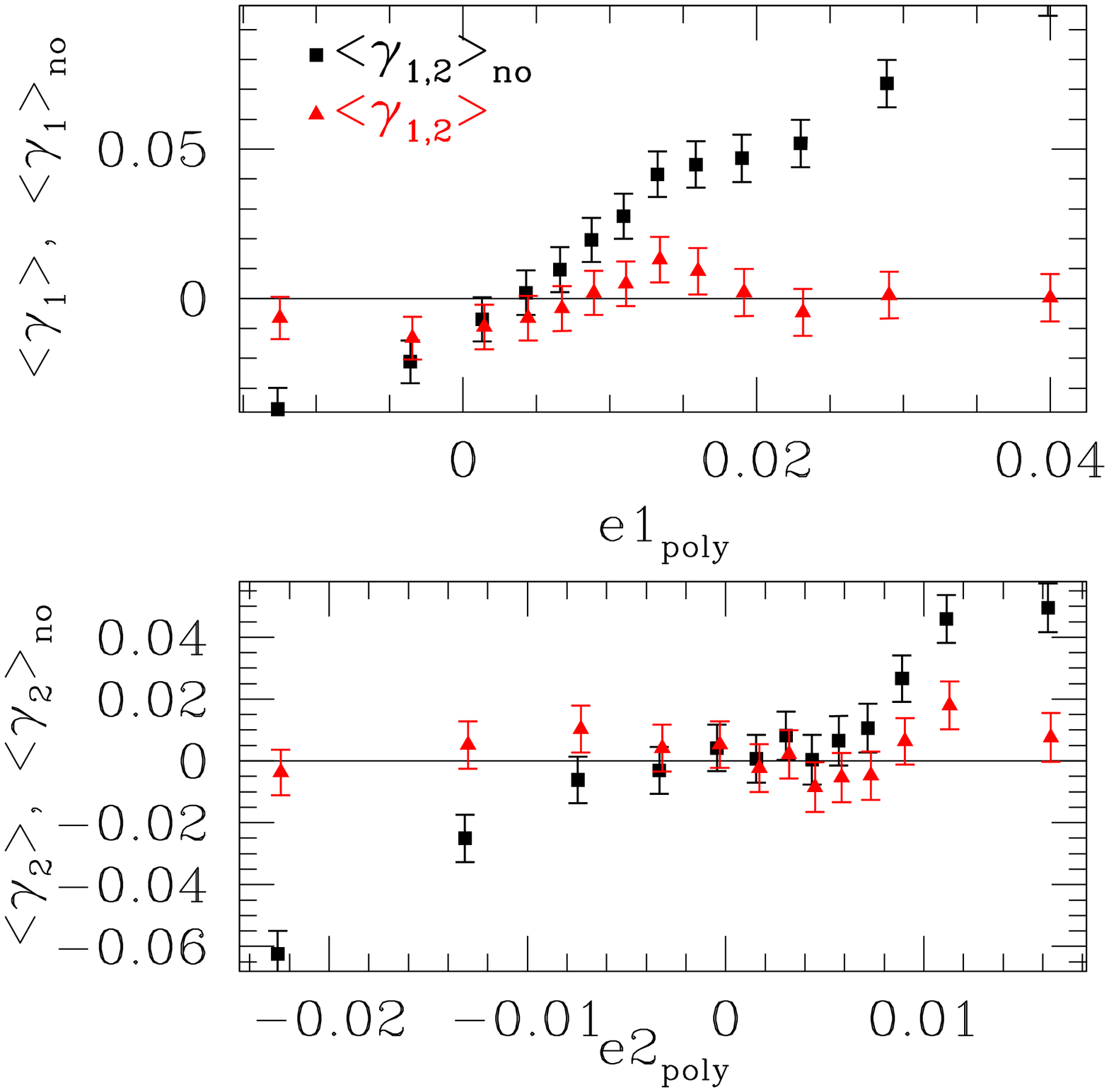}
\includegraphics[scale=0.32]{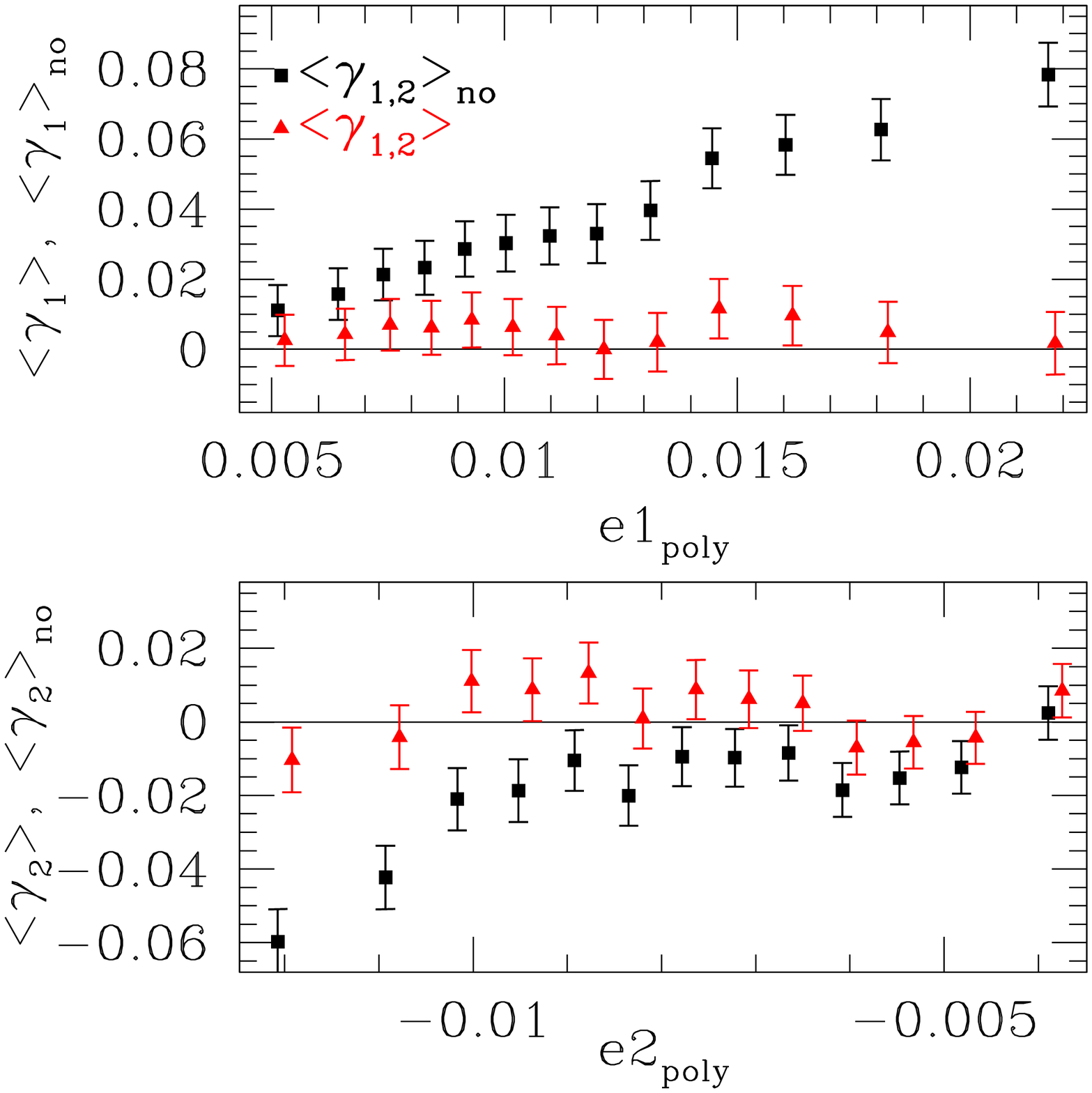}
\caption{
Averaged shear components versus the ellipticity components of the PSF at the galaxy position (polynomial value) with and without anisotropy correction.
{\bf Upper two panels: }AM1-field; {\bf lower two panels: }C04p1-field.
The C04p1-field shows still a small bias ($\gamma_1$ and $\gamma_2$ are on average larger than zero) after the anisotropy correction.
}
\label{fig:ecorrpol_shear}
\end{figure}

\begin{figure}
\centering
\includegraphics[scale=0.4]{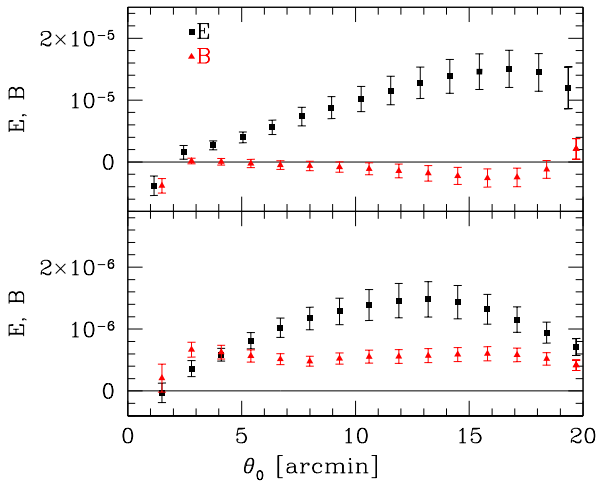}
\caption{E and B-modes of uncorrected ({\bf upper panel}) and anisotropy corrected stars ({\bf lower panel}).
The signal of the corrected stars is significantly inconsistent with zero.
Note, however, the cosmic shear signal is about 10 times larger.
}
\label{fig:Sterne_ani}
\end{figure}

\paragraph{I. Shear - PSF anisotropy.  }
We calculate the average 
\be
\langle e_{1,2}^*\, \gamma_{1,2} \rangle=\frac{1}{N}\sum_{i=1}^N e_{1,2}^{*,{\rm pol}}(\vek{\theta}_i)\,\gamma_{1,2}(\vek{\theta}_i)
\label{eq:crossmultiply}
\ee
before and after anisotropy correction.
The quantities $\gamma_{1,2}$ are the shear components of the galaxies and $e_{1,2}^{*,{\rm pol}}$ are the components of the polynomial fit function (see Sect. \ref{sec:catalogue}) at the position $\vek{\theta}_i$ of the galaxy.
The result is displayed in Fig. \ref{fig:anitest}.
For some data sets the average, $\langle e_i^*\, \gamma_i \rangle$, is not consistent with zero after anisotropy correction.
However, the average is reduced by at least a factor of 20 by applying the anisotropy correction.
%
%
%
\paragraph{II. Average shear.  }
We calculate the average shear $\langle \gamma_{1,2} \rangle$ of all galaxies in the magnitude bin $R \in [21.5,24.5]$ for each GaBoDS-field, before and after anisotropy correction, see Fig. \ref{fig:ani_shear}.
For most of the fields the average shear values, $\langle \gamma_{1,2}\rangle$, are significantly different from zero before the anisotropy correction and $\gamma_1$ is on average significantly larger than zero.
The scatter is strongly reduced after anisotropy correction.
For most of the fields the average shear is consistent with zero.
%
%
%
\paragraph{III. Average shear in bins.  }
A further simple test of residual systematics is given in Fig. \ref{fig:ecorrpol_shear}, where we display for two fields the average shear as a function of the PSF-anisotropy (value of the polynomial fit function at the galaxy position).
The two examples in Fig. \ref{fig:ecorrpol_shear} show that the anisotropy correction works well for the AM1-field, whereas the C04p1-field shows still a small bias in both shear components (the average shear values are on average larger than zero) after the anisotropy correction (see table \ref{SumTab} for details of individual fields).
%
%
\paragraph{IV. E- and B-modes of stars.  }
In order to estimate the amount of residual systematics in the data set, we utilise the stars used for the anisotropy correction. 
Therefore the correlation functions $\xi_\pm$ are measured for all stars before and after the anisotropy correction and the corresponding E- and B-modes ($\langle M_{\rm ap}^2\rangle$ and $\langle M_{\perp}^2\rangle$) are calculated.
The results are presented in Fig. \ref{fig:Sterne_ani}.
After the anisotropy correction we still measure a significant signal which is, however, ten times lower than before correction and ten times lower than the expected cosmic shear signal.

Notable is the significantly larger E- than B-mode signal of uncorrected stars indicating that on average the PSF-anisotropy pattern influences the E-mode signal of galaxies more strongly than the B-mode signal.
%
%
%
\subsection{Cross-correlation of galaxies and stars}
\subsubsection{$\xi_{\rm SYS}^+$}
A way to estimate the amount of systematics left in the cosmic shear signal (like an insufficient PSF anisotropy correction) is given by the cross-correlation of uncorrected ellipticities of stars and PSF-corrected ellipticities of galaxies \citep{btb03,bmr03}.
Since we use the correlation functions $\xi_\pm$ to estimate cosmological parameters we propose a similar estimator:
\be
\xi^+_{\rm SYS}(\theta)\equiv \sum_{i=1}^{N_{\rm f}}\frac{\langle \epsilon^* \gamma \rangle_i^+(\theta) \, |\langle \epsilon^* \gamma \rangle_i^+(\theta)|}{\langle |\epsilon^*|^2\rangle_i(\theta)}\,W_i(\theta),
\label{eq:crosscorr}
\ee
where $N_{\rm f}$ is the total number of fields, $\langle |\epsilon^*|^2\rangle_i$ is the auto-correlation function of stars and $W_i$ is the statistical weight of a single field obtained by bootstrapping on galaxy basis, see Sect. \ref{para:boot1}. 
The cross-correlation function is given by
\be
\langle \epsilon^* \gamma \rangle^\pm(\theta)=\frac{\sum_{i,j}^N w_i^* w_j^{\rm gal} (\epsilon^*_{{\rm t},i}\gamma_{{\rm t},j} \pm \epsilon^*_{\times,i}\gamma_{\times,j})\Delta_{ij}(\theta)}{\sum_{i,j}^N w_i^* w_j^{\rm gal} \Delta_{ij}(\theta)},
\label{eq:crosscorr2}
\ee
where we set $w_i^*=1$ and calculate $w_j^{\rm gal}$ with Eq. (\ref{eq:weight}).
The quantity $\epsilon^*$ is the uncorrected stellar ellipticity and $\gamma$ is the shear estimate of the galaxies.

The definition of $\xi_{\rm SYS}^+$ takes into account that the cosmic shear signal, $\xi_+$, can be negative for individual fields.
In combination with the normalisation of the auto-correlation function of stars, $\xi_{\rm SYS}^+$ is directly comparable to the signal, $\xi_+$.
The results are displayed in Fig. \ref{fig:xi_plot}.
The signal $\xi_+$ is about 10 times larger than the cross-correlation signal $\xi_{\rm SYS}^+$.
We conclude that our way to correct for the PSF anisotropy is sufficient to obtain a clean cosmic shear signal using the two-point shear correlation function.
\subsubsection{$M_{\rm cross;E}$ and $M_{\rm cross;B}$}
We calculate the two-point cross-correlation function $\langle \epsilon^* \gamma \rangle^\pm$ for every field using Eq. (\ref{eq:crosscorr2}).
The cross-correlation function $\langle \epsilon^* \gamma \rangle^\pm$ is inserted in Eq. (\ref{eq:mapdis}) and Eq. (\ref{eq:Bmapdis}) to obtain 
\be
M_{\rm cross;E}\!\equiv\!\langle M_{\rm ap}^* M_{\rm ap}^{\rm gal} \rangle(\theta)\!=\!\frac{\sum_{i=1}^{N_{\rm f}}\langle M_{\rm ap}^* M_{\rm ap}^{\rm gal} \rangle_i(\theta)\,W_i(\theta)}{\sum_{i=1}^{N_{\rm f}}W_i(\theta)}
\ee
and 
\be
M_{\rm cross;B}\!\equiv\!\langle M_\times^* M_\times^{\rm gal} \rangle(\theta)\!=\!\frac{\sum_{i=1}^{N_{\rm f}}\langle M_\times^* M_\times^{\rm gal} \rangle_i(\theta)\,W_i(\theta)}{\sum_{i=1}^{N_{\rm f}}W_i(\theta)},
\ee
respectively.
The quantity $N_{\rm f}$ is the total number of fields and $W_i$ is the statistical weight of a single field obtained by bootstrapping on galaxy basis, see Sect. \ref{para:boot1}.  
The results are displayed in Fig. \ref{fig:GaBoDS_single} and \ref{fig:GaBoDS_EB} for the different data sets and for the total data set, respectively.
The cross-correlation signal, $M_{\rm cross;E}$, is significantly non-zero for the C0-fields and shows the same characteristic as the E-mode signal.
Hence our anisotropy correction seems to work only insufficiently for these fields.
Note that, in contrast to \citet{smv06}, a function similar to $\xi_{\rm SYS}^+$ (a function normalised with the auto-correlation of stars, $\langle |M^*|^2\rangle$) is not calculated for $M_{\rm ap}$.
This is because in our case we analyse single fields which are shallower and smaller in size, hence the $M_{\rm ap}$ auto-correlation function of stars is very noisy and can have zero-crossings.
%
%
%
%
\section{Cosmic shear analysis}
\label{sect:cosmicshearanalysis}
\subsection{Error estimates and field weights}
\paragraph{Bootstrapping on galaxy basis.  }
\label{para:boot1}
To obtain the statistical weights of $\langle M_{\rm ap}^2\rangle$ and $\xi_\pm$ for each field $i$ and angular bin $j$ we make 200 bootstrap samples of the galaxy catalogue as follows.
We randomly draw $N_{\rm gal}$ times galaxies from the galaxy catalogue for each field $i$ with putting back (where $N_{\rm gal}$ is the total number of galaxies in the field $i$) and place them at the same position with the same orientation as before.
For each bootstrap sample, $\langle M_{\rm ap}^2\rangle_{ij}$ and $\xi_{\pm,ij}$ are calculated for each angular bin $j$.
The statistical error for each field and bin is then the bootstrapping variance $\sigma_{ij}^2$. 

The average measurement signal of the total survey in each bin $j$ is calculated as
\be
\langle M_{\rm ap}^2\rangle_j= \frac{\sum_{i=1}^{N_{\rm f}} \langle M_{\rm ap}^2\rangle_{ij}\, W_{ij}}{\sum_{i=1}^{N_{\rm f}} W_{ij}} \, \quad {\rm and}
\label{eq:averagesignal}
\ee
\be
\xi_{\pm;j}= \frac{\sum_{i=1}^{N_{\rm f}} \xi_{\pm;ij}\, W_{ij}}{\sum_{i=1}^{N_{\rm f}} W_{ij}},
\label{eq:averagesignal2}
\ee
where $N_{\rm f}$ is the total number of fields.
As weight $W_{ij}$ we use the reciprocal bootstrapping variance $W_{ij}=1/\sigma_{ij}^2$.
In Fig. \ref{fig:weight} (App. \ref{appendixa}) the relative statistical weights of the GaBoDS fields calculated for $\langle M_{\rm ap}^2\rangle$ versus the angular bin (aperture radius $\theta_0$) are displayed.
The statistical weights of fields with a low number density of galaxies are lower than of those with a high number density, as expected.
This can also be seen in Fig. \ref{fig:singlefields} (upper and lower left panels) where the square root of the variance, $\sigma_{ij}$ is plotted for two fields with different number densities (FDF-field: $1.9\times 10^4$ galaxies and C04p3-field: $1.3\times 10^4$ galaxies).
Clearly visible are also the large error bars for small angular scales since the number of pairs is small for them.
\paragraph{Bootstrapping on field basis.  }
\label{para:boot2}
The measurement error of the combined signal for each angular bin $j$ is obtained by 2000 bootstrap samples of the field sample as follows.
We randomly draw $N_{\rm f}$ times fields out of the field sample with putting back.
This bootstrap sample is combined according to Eqs. (\ref{eq:averagesignal}) and (\ref{eq:averagesignal2}).
With the 2000 bootstrap samples we estimate the PDF of the statistical errors in the combined signal including cosmic variance and obtain covariances of the errors for the final analyses.
With the PDF we obtain for each angular bin $j$ the error of the combined signal by calculating the $68\,\%$ confidence intervals about the mean.
For equal weights the calculated $1\,\sigma$ variance in each bin is comparable to the usual field-to-field variance in each bin as the PDF of the bootstrap sample is almost symmetric in our case. 
%
%
%
\begin{figure*}
\centering
\resizebox{\hsize}{!}{\includegraphics[width=\textwidth,clip]{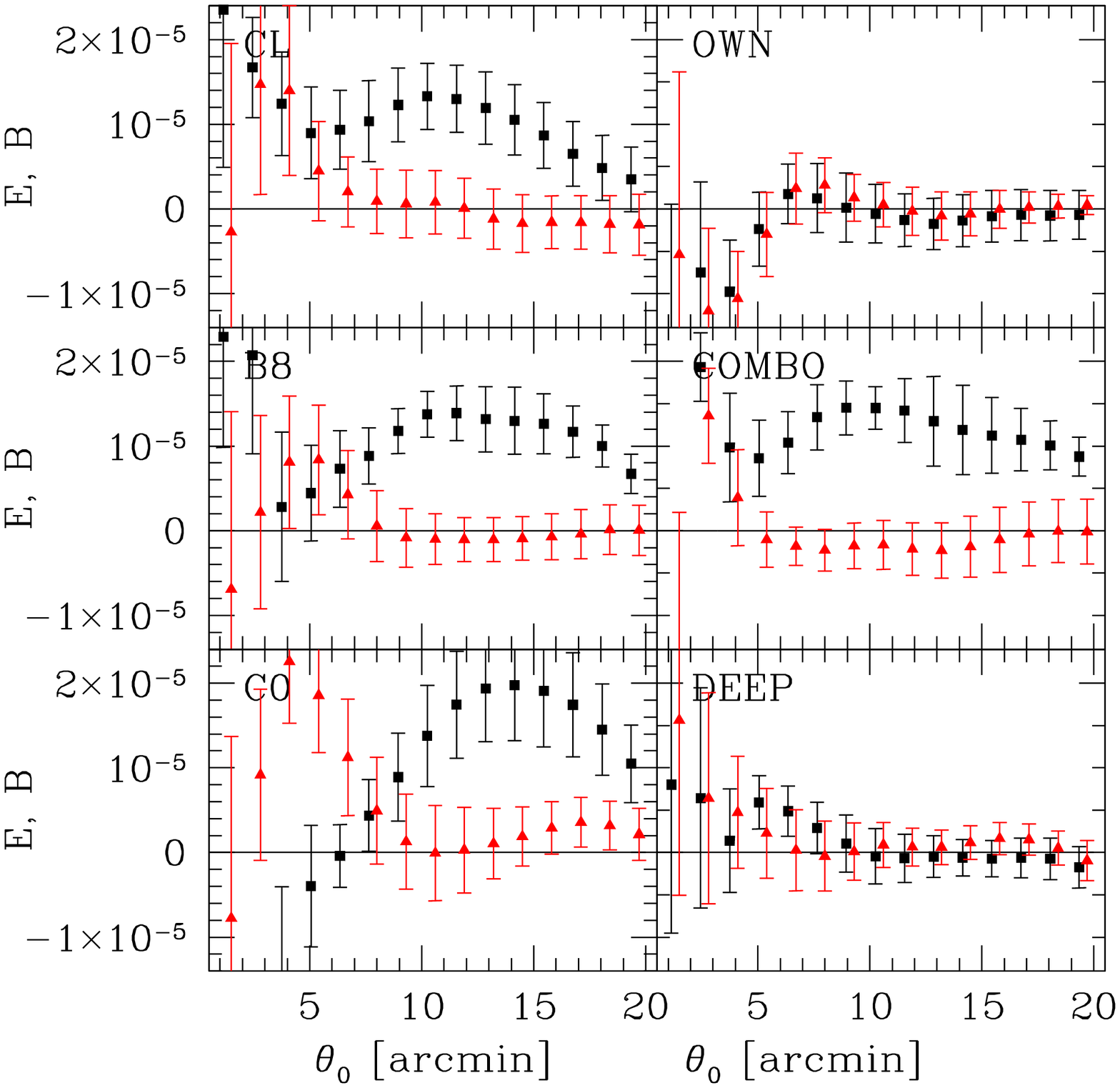}
\includegraphics[width=\textwidth,clip]{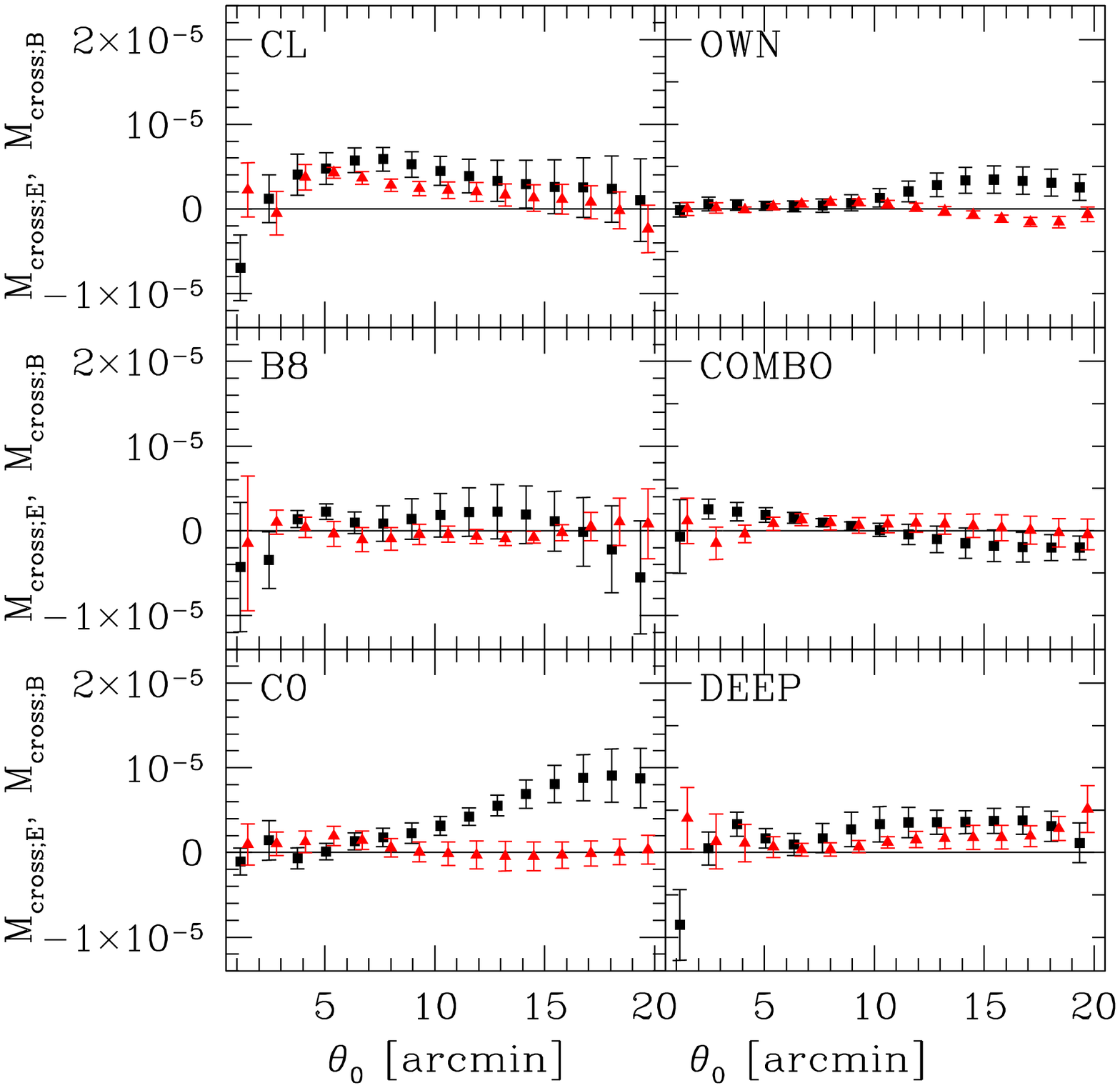}
}
\caption{Measurements of the aperture mass statistics in $R$-band for all six data sets.
{\bf Left panel : } E-mode (squares) and B-mode (triangles).
{\bf Right panel: } $M_{\rm cross; E}$ (squares) and $M_{\rm cross; B}$ (triangles).
For clarity E- and B-mode points are slightly offset.
}
\label{fig:GaBoDS_single}
\end{figure*}
\subsection{Cosmic shear signal }
\label{sec:signal}
A powerful way to reveal possible systematic errors is the application of the aperture mass statistics as it provides an unambiguous splitting of E- and B-modes.
The presence of non-vanishing B-modes is a good indicator for systematics arising, for instance, from an imperfect anisotropy correction.
However, \citet{kse06} found in their work a mixing of E- and B-modes if a cut-off in $\xi_\pm$ on small angular scales occurs.
According to \citet{kse06} the deviation of the biased E- and B-mode (the E- and B-mode signals are smaller due to the cut-off) from the true E- and B-mode signal is only $1\,\%$ for angular scales up to $4\,{\rm arcmin}$ for a cut-off in $\xi_\pm$ at $6\,{\rm arcsec}$. 
In our work we calculate $\langle M_{\rm ap}^2\rangle$ from $\xi_\pm$ in the interval $[6^{\prime\prime},40^\prime]$.
For the parameter estimation we only consider the E-mode signal from angular scales larger than $4\,{\rm arcmin}$ (see below), so the bias described is negligible.
\citet{sck06} overcome the general problem and introduce the ring-statistics which unambiguously splits E- and B-modes.

At this stage it is worth mentioning that B-modes can also arise from the intrinsic alignment of galaxies \citep[e.g.][]{hrh00,cnp01} which could be removed, in principle, by using redshift information \citep{kis02,kis03,heh03}.
A source of negative B-modes is the intrinsic shape-shear correlation predicted by \citet{his04} and analysed with numerical simulations by \citet{hwh06}.
In addition, B-modes can arise from clustering of galaxies from which the shear is measured \citep{svm02}.
All three sources of contaminations of the B-mode signal combined are probably not significant as the statistical errors are still quite large.
In Sect. \ref{sec:errors} we discuss the contamination of the E-mode signal by the intrinsic alignment and the intrinsic shape-shear correlation of galaxies which can be much larger and significant.

As pointed out in Sect. \ref{sect:data} we split our data set into six different sets depending on the data source.
The data is heterogeneous with respect to the seeing conditions, observing strategy and other characteristics (see Sect. \ref{sec:dataset}).
Thus splitting of the data into different sets allows us to reveal potential systematics depending on the data source.
In Fig. \ref{fig:GaBoDS_single} the average E- and B-mode signal and the average signal of the cross-correlation between uncorrected stars and PSF corrected galaxies for E- and B-modes, $M_{\rm cross;E}$ and $M_{\rm cross;B}$, of the different data sets are displayed.
In particular the cross-correlation signal, $M_{\rm cross;E}$, of the C0-fields is significantly non-zero and shows the same characteristic as the E-mode signal.
In addition, the B-mode signal within the interval \mbox{$\theta_0 \in [2^\prime,7^\prime]$} is significantly larger than zero, hence we exclude those fields from our final analysis.

\begin{figure}
\centering
\resizebox{\hsize}{!}{\includegraphics[width=\textwidth,clip]{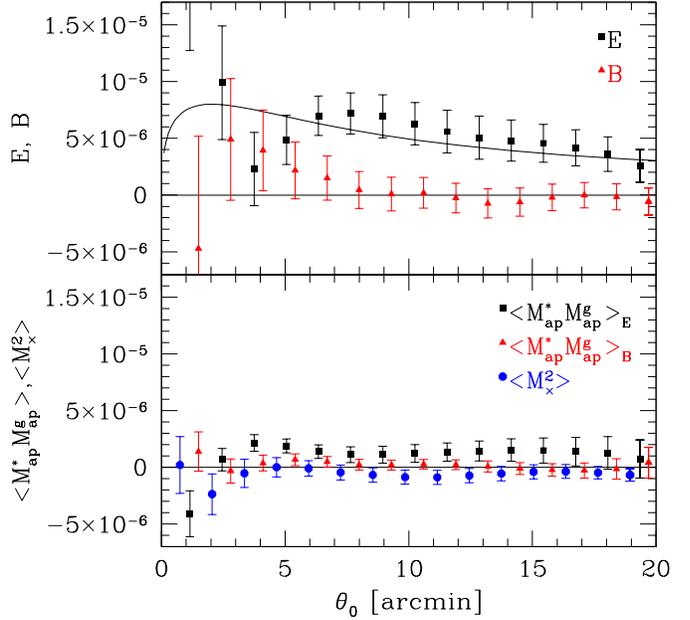}}
\caption{Various measurements of the aperture mass statistics of 52 GaBoDS fields (the C0 fields and CAPO are excluded, see text) in $R$-band ($\approx 13\,{\rm deg}^2$). 
All galaxies in the magnitude range $R \in[21.5,24.5]$ are used for the cosmic shear analysis.
The total number of galaxies is $\approx 7.8 \times 10^5$ corresponding to an effective number density of $n_{\rm eff}=12.5\,{\rm arcmin}^{-2}$.
{\bf Upper panel: } E- and B-mode decomposition of the $M_{\rm ap}$-signal.
The line is a $\Lambda$CDM prediction assuming $\Omega_{\rm m}=0.3$, $\Omega_\Lambda=0.7$, $\Gamma=0.21$ and $\sigma_8=0.9$; the redshift distribution function is given by Eq. (\ref{redshiftdistribution}).
For clarity the B-mode points are slightly offset to the right.
{\bf Lower panel: } Cross-correlation between uncorrected stars- and corrected galaxies for E- and B-modes and $\langle M_\times^2\rangle$.
For clarity the $\langle M_\times^2\rangle$- and $\langle M_{\rm ap}^* M_{\rm ap}^{\rm gal} \rangle_{\rm B}$-points are slightly offset to the left and right, respectively.
}
\label{fig:GaBoDS_EB}
\end{figure}

By excluding the C0-fields we obtain, in the magnitude range $R\in[21.5,24.5]$, $7.8 \times 10^5$ galaxies, corresponding to a number density of $n=16\,{\rm arcmin}^{-2}$.
Taking into account the weights of individual galaxies this results in an effective number density of source galaxies of $n_{\rm eff}=12.5\,{\rm arcmin}^{-2}$. 
In Fig. \ref{fig:GaBoDS_EB} the average E-mode, B-mode, $\langle M_\times^2\rangle$-signal and the average signal of the cross-correlation between uncorrected stars and PSF corrected galaxies of all fields, except the excluded C0-fields, are displayed.
The average B-mode signal is consistent with zero within the $1\,\sigma$-range for \mbox{$\theta_0 > 4\,{\rm arcmin}$}, and the cross-correlation between uncorrected stars and corrected galaxies, $M_{\rm cross;B}$, is consistent with zero.
Hence the B-mode signal does not suffer from an imperfect anisotropy correction.
In addition, the $\langle M_\times^2\rangle$-signal is consistent with zero indicating a clean data set, free from any systematic errors.
The cross-correlation signal, $M_{\rm cross;E}$, however, is about $1\,\sigma$ to $2\,\sigma$ larger than zero for \mbox{$\theta_0>5\,{\rm arcmin}$} and $3\,\sigma$ for $\theta_0 \in [3^\prime,5^\prime$].
This suggests that the PSF-anisotropy correction is not perfect and biases the E-mode signal even though the B-modes are consistent with zero.  
However, $M_{\rm cross;E}$ and $M_{\rm cross;B}$ are not normalised with the auto-correlation function of stars, so $M_{\rm cross;E}$ is not directly comparable with the lensing signal itself. 
Therefore it is hard to judge how large the possible impact of systematics on the E-mode really is.

Taking into account the B-modes, the $\langle M_\times^2\rangle$ and the cross-correlation signals $M_{\rm cross;E}$ and $M_{\rm cross;B}$ we conclude that the influence of systematics on the calculated E-mode signal is negligible on angular scales larger than four arcminutes within the currently attainable accuracy.
For scales below four arcminutes the B-modes are slightly positive (about $1\,\sigma$) and the $M_{\rm cross;E}$-signal has its maximum.
For this reason, we exclude in our further $\langle M_{\rm ap}^2\rangle$-analysis, scales below four arcminutes. 

\begin{figure}
\centering
\resizebox{\hsize}{!}{\includegraphics[width=\textwidth,clip]{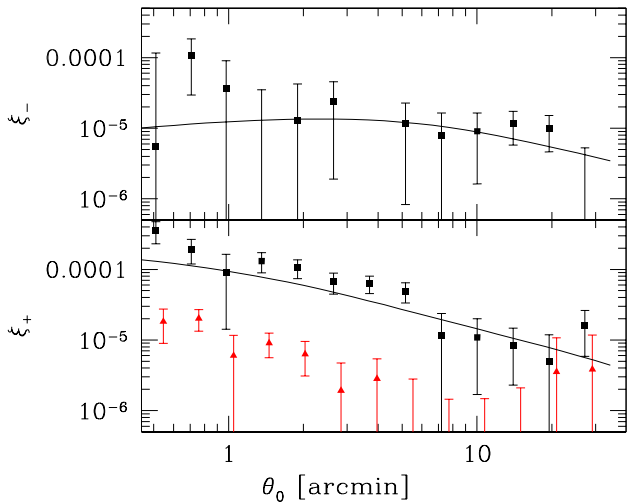}}
\caption{Measured two-point correlation functions $\xi_+$ and $\xi_-$.
In the lower panel $\xi_{\rm SYS}^+$ (see Eq. \ref{eq:crosscorr}) is plotted additionally (triangles).
The line is a $\Lambda$CDM prediction (same parameters as in Fig. \ref{fig:GaBoDS_EB}).  
}
\label{fig:xi_plot}
\end{figure}

In Fig. \ref{fig:xi_plot} the average $\xi_-$-, $\xi_+$- and $\xi_{\rm SYS}^+$-signals of all fields, except the excluded C0-fields, are plotted.
The $\xi_{\rm SYS}^+$-signal is at least a factor of 10 smaller than the $\xi_+$-signal so we conclude that the two-point shear correlation function is not significantly influenced by an imperfect anisotropy correction.

Both, the aperture mass dispersion (in the interval $\theta_0 \in [4^\prime,20^\prime]$) and the two-point shear correlation function (in the interval $\theta_0 \in [0\myarcmin 8,33^\prime])$ are used in the following to estimate cosmological parameters.
%
%
%
%
\subsection{Photometric redshifts}
The GaBoDS data set includes $2\,{\rm deg}^2$ of deep {\it UBVRI}-band observations from the DPS, which yields photometric redshift information for about $8\%$ of the objects considered for the cosmic shear analysis.
The redshift catalogue was obtained using the {\tt hyper-z} public software developed by \citet{bmp00}.
For the photometric redshift estimation we used the observed galaxy templates by \citet{cww80}.

The synthetic galaxy templates derived from the library of \citet{brc93} yield a smaller scatter in the difference between photometric and spectroscopic redshifts for galaxies with magnitudes $R<23$.
However, the redshift distribution of galaxies in the magnitude interval $R\in [21.5,24.5]$, obtained with the galaxy templates by \citet{cww80}, matches the redshift distribution obtained from the GOODS-MUSIC sample \citep{gfd06} slightly better than that obtained with galaxy templates by \citet{brc93}.

For the Chandra Deep Field South (name in our paper: AXAF), a database of spectroscopically observed objects from the VIMOS-VLT-Deep-Survey \citep[VVDS,][]{fvp04} exists.
A comparison between the redshifts of 500 VVDS-objects and photometric redshifts obtained with {\tt hyper-z}, in the considered magnitude interval $R\in [21.5,24.5]$, reveals a systematic underestimation of the photometric redshifts of $0.055\,(1+\langle z\rangle)$.
This is corrected for in an ad-hoc manner by adding $0.055\,(1+\langle z\rangle)$ to each photo-$z$ of a galaxy.

\begin{figure}
\centering
\resizebox{\hsize}{!}{\includegraphics[width=\textwidth,clip]{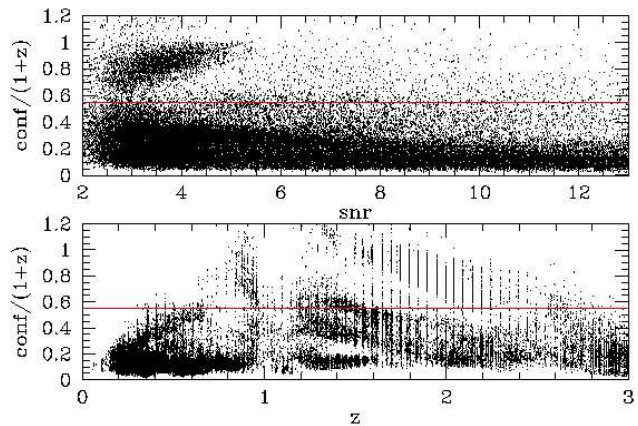}}
\caption{{\bf Upper panel: } Relative {\tt hyper-z} $68\,\%$-confidence interval, conf/(1+z), of lensing galaxies and its dependence on signal-to-noise ratio (snr) in $R$-band.
{\bf Lower panel: } Relative confidence interval and its dependence on photometric redshift $z$.
The vertical line indicates the cut described in the text.
The photometric redshifts are obtained from 7 WFI fields of the DPS (with 5 broadband filters: {\it UBVRI}) in the magnitude interval $R \in [21.5,24.5]$.
}
\label{fig:test_conf}
\end{figure}

To estimate the redshift distribution of the background sources we associate the redshift catalogue with the lensing catalogue.
In this way the weighting of individual galaxies (see Eq. \ref{eq:weight}) is taken into account in the estimation of the redshift distribution. 
We only consider those galaxies with photometric redshift estimates in the interval $z\in [0,3]$.
In addition, we only take those photo-z estimates into account with a {\tt hyper-z} $68\,\%$-confidence interval smaller than $0.55\,(1+z)$.
With this cut we exclude objects with highly uncertain redshift estimates which are also galaxies with a low signal-to-noise ratio in $R$-band, see Fig. \ref{fig:test_conf}.
%
%

We obtain $6.2 \times 10^4$ objects with accurate photometric redshifts corresponding to $70\,\%$ of the DPS-lensing objects and about $8\,\%$ of all lensing objects.
With these objects we estimate a redshift distribution of the DPS-lensing galaxies in the given magnitude bin. 
To acquire a smooth redshift distribution for all galaxies of the GaBoDS galaxy lensing catalogue we fit the following distribution function $p_{\rm fit}$, introduced by \citet{blp06}, to the measured redshift distribution,
\be
p_{\rm fit}(z)=A\,\left[p_1(z)\,H(z_{\rm t}-z)+p_2(z)\,H(z-z_{\rm t}
)\right], 
\label{redshiftdistribution}
\ee
where $H$ denotes the Heaviside step function, 
\be
p_1(z) = \left(\frac{z}{z_0}\right)^\alpha \exp\left[-\left(\frac{z}{z_0}\right)^\beta\right]
\ee
and
\be
p_2(z) = \exp{\left[\left(\frac{z_t}{z_1}\right)^\gamma - \left(\frac{z}{z_1}\right)^\gamma \right]}\,p_1(z_{\rm t});
\ee
the normalisation $A$ is obtained by
\be
\int_0^{z_{\rm t}} \dd z\, p_1(z) + \int_{z_{\rm t}}^3 \dd z\, p_2(z) = 1. 
\ee
The fit function was binned in the same way as the measured distribution. 
For $z_{\rm t}=1$ we minimise $\chi^2$ and obtain as best-fit parameters $z_0=0.27\pm 0.14$, $\alpha=2.84\pm 0.96$, $\beta=1.40 \pm 0.37$, $\gamma=2.34 \pm 0.53$ and $z_1=2.16 \pm 0.22$ which results in a mean redshift of $\bar z=0.78$.
The given errors are $1\,\sigma$ statistical uncertainties.
%
%
\begin{figure}
\centering
\resizebox{\hsize}{!}{\includegraphics[width=\textwidth,clip]{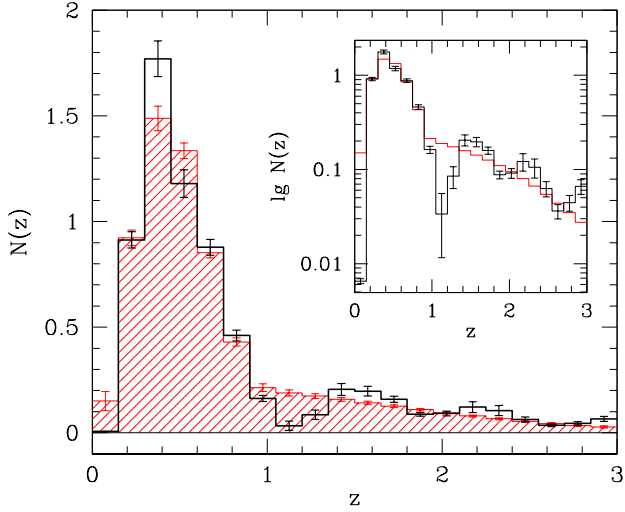}}
\caption{Normalised number density of galaxies as a function of photometric redshift.
Solid histogram: measured redshifts, shaded histogram: fit given by Eq. (\ref{redshiftdistribution}).
The photometric redshifts are obtained from 7 WFI fields of the DPS (with 5 broadband filters: {\it UBVRI}) for $6.2 \times 10^4$ objects ($\approx 70\%$ of the lensing objects) in the magnitude interval $R \in [21.5,24.5]$.
To better display the tail of the distribution we plot inside the diagram a logarithmic plot of $N(z)$.
The error bars are the field-to-field variance of the 7 WFI fields.
}
\label{fig:redshifts}
\end{figure}

In Fig. \ref{fig:redshifts} the observed and fitted redshift distributions of the DPS galaxy sub-sample in the magnitude interval $R \in [21.5,24.5]$ are displayed. 
Note the dip at $z=1.1$ in the observed redshift distribution.
This dip is present in the estimated redshift distributions of each DPS field, yet not for the GOODS-MUSIC sample which comprises infrared bands.
We assume that the dip is probably a systematic effect due to the lack of infrared information.
We therefore downweighted this redshift bin for the fit.

A more proper way to determine the redshift distribution from the photometric redshift estimates would be to deconvolve the latter with the width of the phot-z uncertainties.
However, the corresponding change in the ``true'' redshift distribution would be fairly small, yielding a correction factor which is much smaller than the statistical uncertainties of the shear measurements. 
We have therefore not attempted such a deconvolution and do not consider this negligible bias of the redshift distribution in the following error estimation of $\sigma_8$ (Sect. \ref{sec:errors}). 
\subsection{Parameter estimation}
The measured cosmic shear signal is related to the dark matter density power spectrum, which in turn depends on cosmological parameters and the source redshift distribution.
To estimate the parameters \mbox{$\vek{\alpha}=(z_0,z_1,\alpha,\beta,\gamma,\sigma_8,\Omega_{\rm m},h,\Omega_\Lambda)$} from our measured data vector $\vek{d}$ (either the binned $\langle M_{\rm ap}^2\rangle$- or $\xi_\pm$-values) we use the generalised likelihood. 
The probability distribution function (PDF) of the model parameters $\vek{\alpha}$ given the $p$-dimensional data vector $\vek{d}$ is called the posterior PDF $P(\vek{\alpha}|\vek{d})$.
It is obtained using Bayes theorem as follows
\be
P(\vek{\alpha}|\vek{d})\propto P_{\rm prior}(\vek{\alpha})P(\vek{d}|\vek{\alpha}),
\label{eq:bayes}
\ee
where $P(\vek{d}|\vek{\alpha})$ is the usual likelihood function, $P_{\rm prior}(\vek{\alpha})$ is a prior of $\vek{\alpha}$ and $\vek{m}(\vek{\alpha})$ is the $p$-dimensional model vector.
Assuming that the noise residual \mbox{$\vek{n}=\vek{d}-\vek{m}$} obeys Gaussian statistics the log-likelihood function reads
\be
\ln P(\vek{d}|\vek{\alpha})=-\frac{1}{2}(\vek{d}-\vek{m})^t C^{-1} (\vek{d}-\vek{m})+{\rm const}.
\ee
An unbiased estimator of the inverse covariance matrix is given by
\be
 C^{-1}=\frac{N_{\rm f}-p-2}{N_{\rm f}-1}\tilde C^{-1},
\ee
where $p$ is the size of the data vector (in our case the number of angular bins, with \mbox{$p=15$}), $N_{\rm f}$ is the number of statistical independent data vectors (in our case the number of fields, \mbox{$N_{\rm f}=52$}). 
Without the factor, \mbox{$(N_{\rm f}-p-2)/(N_{\rm f}-1)$}, the error contours would be underestimated and strongly depend on the number of bins, $p$.
For a detailed discussion of how to obtain unbiased estimates of inverse covariance matrices, we refer to Hartlap et al. (in preparation). 
The quantity
\be
 \tilde C_{ij}=\frac{1}{N_{\rm boot-1}}\sum_{l=1}^{N_{\rm boot}}(d_i^{(l)}-\bar d_i) (d_j^{(l)}-\bar d_j)
\ee
is the estimated noise covariance matrix with the angular bins $i,j\in \{1,2,...,p\}$.
The quantity $\bar d$ is the average signal of all fields in the considered bin, see Eqs. (\ref{eq:averagesignal}) and (\ref{eq:averagesignal2}), and $N_{\rm boot}=2000$ is the number of bootstrap samples on field basis (Sect. \ref{para:boot2}).

In many other works on cosmic shear the covariance matrix is computed analytically (especially for those cosmic shear surveys consisting of a large contiguous field).
The covariance matrix is decomposed as \mbox{$\tilde C=C_{\rm n}+C_{\rm s}$}, where $C_{\rm n}$ is the statistical noise and $C_{\rm s}$ is the cosmic variance covariance matrix.
The cosmic variance covariance matrix is then computed according to \citet{svk02} assuming a fiducial cosmological model and an effective survey area.  
Note that in contrast to this approach we estimate our covariance matrix {\it directly from the data} without any further assumptions.
This is possible as our fields are widely separated in the sky; hence they are statistically independent.

To calculate the model vector $\vek{m}(\vek{\alpha})$, we have to specify the dark matter density power spectrum.
Our measurements probe scales which are affected by the non-linear growth of structure.
We calculate the non-linear power spectrum using the fitting formula proposed by \citet{ped96}. 
The fitting formula is based on a semi-analytical prescription for the evolution of the dark matter power spectrum.
Although this formula is relatively simple it reproduces the main features of the standard cosmological model and is accurate enough for our purposes, considering the large statistical errors of our measurements.
To estimate cosmological parameters we assumed two simple $\Lambda$CDM models:
\begin{itemize}
\item {\bf A}: a flat universe: $\Omega_{\rm m}+\Omega_\Lambda=1$ and
\item {\bf B}: a $\Lambda$-universe with $\Omega_\Lambda$, $\Omega_{\rm m}\in[0,1.5]$.
\end{itemize}
For both models we assumed a negligible baryon content \citep[the shape parameter is given by $\Gamma=\Omega_{\rm m}\,h$ and the transfer function is given in ][]{bar86} and a primordial spectral index of $n=1$.
The remaining free parameters in our following likelihood analyses are therefore:
\begin{itemize}
\item the five fit parameters of the redshift distribution ($z_0,z_1,\alpha,\beta,\gamma$) with the strong constraint from our observed redshift distribution,
\item Hubble constant with a strong constraint of \mbox{$h=0.7 \pm 0.07$} (supported by the HST key project),
\item mass power spectrum normalisation $\sigma_8$,
\item matter density with the weak constraint $\Omega_{\rm m} \geq 0$,
\item in addition for model {\bf B}: $\Omega_\Lambda$.
\end{itemize}
To apply Eq. (\ref{eq:bayes}) we have to specify the probability function of the priors, $P_{\rm prior}(\vek{\alpha})$.
It is given by a multivariate Gaussian distribution
\be
P_{\rm prior}(\vek{\alpha})\propto \exp{\left\{-\frac{1}{2}(\vek{\alpha}-\vek{\bar \alpha})^t C_{\rm prior}^{-1}(\vek{\alpha}- \vek{\bar \alpha})\right\}}.
\ee
The correlation between the five fit parameters of the redshift distribution is taken into account in the prior covariance matrix, $C_{\rm prior}$.
\subsubsection{MCMC}
We do not calculate the likelihood function on a grid.
It would be eight- or nine-dimensional in our case ($\sigma_8,\Omega_{\rm m},h$ or $\sigma_8,\Omega_{\rm m},h,\Omega_\Lambda$, depending on model {\bf A} or {\bf B}, and the five fit parameter of the redshift distribution) and would therefore be too time consuming.
The Monte Carlo Markov Chain (MCMC) technique overcomes this computational limitation.
The likelihood function is no longer calculated at fixed positions but at randomly selected positions along the Markov Chain.

The MCMC code in our work is based on the Metropolis-Hastings algorithm, see \citet{mrr53} and \citet{has70}, or \citet{tdv05} for recent implementations to estimate cosmological parameters.
It performs a random walk through the parameter space $S_i = (\alpha_1^{(i)},...,\alpha_{8/9}^{(i)})$ with $i \in [1,N_{\rm s}]$ ($N_{\rm s}$ is the number of accepted steps).
The starting point, $S_0$, can be somewhere in the parameter space.
We choose it to be close to the expected likelihood maximum.
Then a candidate for the next step $S_{i+1}$, using a proposal PDF $Q(S_i,S_{i+1})$, is drawn.
The value $S_{i+1}$ is accepted if the following inequalities are fulfilled, either 
\be
P(S_{i+1}|d)\ge P(S_i|d)\,\, {\rm or}\quad \frac{P(S_{i+1}|d)}{P(S_i|d)} \ge x \frac{Q(S_i,S_{i+1})}{Q(S_{i+1},S_i)},
\label{eq:accept}
\ee
where $x \in [0,1]$ is a random number drawn from a uniform distribution.
If the candidate is accepted we move to $S_{i+1}$ and repeat the procedure.

In general the choice of $Q(S_i,S_{i+1})$ is free.
To achieve convergence in a reasonable time we choose, as in \citet{tdv05}, a multivariate Gaussian distribution
\be
Q(S_i,S_{i+1})\propto \exp{\left\{-\frac{1}{2}[S_i-S_{i+1}]^t M^{-1}[S_i-S_{i+1}]\right\}}.
\label{eq:multigauss}
\ee
For $M$ we choose during a pre-phase (first 1000 steps along the chain) the Fisher matrix $F_{ij}$ at the position $S_0$, the given starting point:
\be
F_{ij}= F_{ij}^{\rm prior}+\sum_{k,l}\frac{\partial m_k}{\partial \alpha_i}[C^{-1}]_{kl}\frac{\partial m_l}{\partial \alpha_j}
\ee
where 
\be
F_{ij}^{\rm prior}=\frac{\partial^2}{\partial \alpha_i \partial \alpha_j} \ln P_{\rm prior}(\vek{\alpha})
\ee
is the Fisher matrix of the priors.
After the pre-phase we replace $M$ by the covariance matrix, ${\rm cov}(\alpha_i,\alpha_j)$, of the parameters estimated from the first 1000 steps.
As $Q$ in Eq. (\ref{eq:multigauss}) satisfies $Q(S_i,S_{i+1})=Q(S_{i+1},S_i)$, Eq. (\ref{eq:accept}) simplifies to $P(S_{i+1}|d)/P(S_i|d) \ge x $.

Usually several independent MCMCs are started.
For every chain the first $S_i$ values are rejected until the MCMCs have forgotten about their initial start position.
In this way a possible bias due to the starting position is avoided.
To estimate the size of the so-called burn-in-phase we use the Gelman \& Rubin test \citep{ger92}.
For each parameter, $\alpha_{kl}$, first the variance of the parameters, $\sigma_{\alpha_{kl}}^2$, inside every chain $k$, and then the average variance \mbox{$\bar \sigma_{\alpha_l}^2=1/k \sum_k \sigma_{\alpha_{kl}}^2$} of all chains is calculated.
Next the average of each parameter, $\bar \alpha_{kl}$, inside each chain, and then the variance of all averages, $\sigma_{\bar \alpha_l}^2$, is calculated.
The average, $\bar \alpha_{kl}$, and variance, $\sigma_{\alpha_{kl}}^2$, inside each chain is calculated using only $S_i$, with $i\in[n,2n]$.
For ${n \to \infty}$ both variances, $\sigma_{\bar \alpha_l}^2$ and $\bar \sigma_{\alpha_l}^2$, converge to the same value.
Therefore the ratio 
\be
R=\frac{\sigma_{\bar \alpha_i}^2}{\bar \sigma_{\alpha_i}^2}+\frac{n}{n-1}
\ee
calculated for each parameter $\alpha_l$ is a reasonable control parameter for the convergence of each chain, see \citet{tdv05}.
As long as $R$ is larger than 1.2 we reject all $S_i$ with $i<n$.

To decorrelate the $S_i$ values a thinning is applied.
Only every fifth value of the chain is used as proposed by \citet{tdv05}.

We marginalise over the Hubble parameter $h$ and the fit parameter of the redshift distribution, $z_0$, $\alpha$, $\beta$, $\gamma$ and $z_1$, meaning that we project all points of the MCMC onto the ($\Omega_{\rm m},\sigma_8$), ($\Omega_\Lambda,\sigma_8$) or ($\Omega_{\rm m},\Omega_\Lambda$)-plane.
Compared to the prior, the Hubble parameter and the redshift distribution parameters cannot be improved with the data at hand.
To determine the ($1\,\sigma, 2\,\sigma$ and $3\,\sigma$) contours, the point number density in the ($\Omega_{\rm m},\sigma_8$), ($\Omega_\Lambda,\sigma_8$) and ($\Omega_{\rm m},\Omega_\Lambda$)-plane is smoothed and the logarithm of the number density map is calculated.
The maximum log-density is subtracted everywhere.
In this way $-\Delta \chi^2/2=\ln{L}-\ln{L_{\rm max}}$ is obtained.
\subsubsection{Results} 
\begin{figure}
\centering
\includegraphics[scale=0.37,angle=-90]{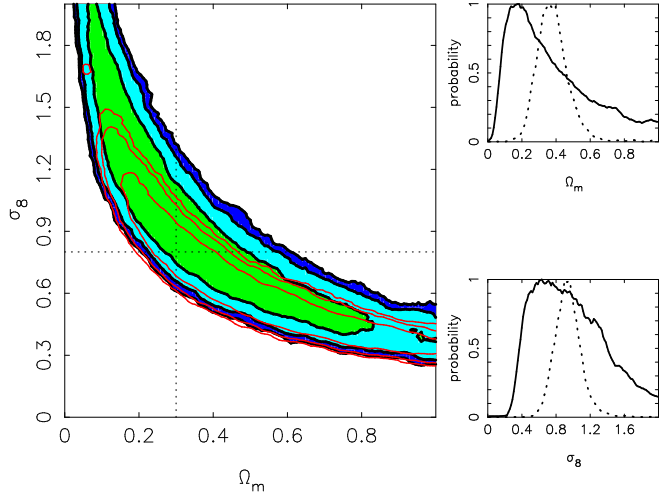}
\caption{
Model {\bf A}: joint constraints on $\Omega_{\rm m}$ and $\sigma_8$ obtained from our cosmic shear analysis of the GaBoDS data set using the two-point shear-correlation function and the Peacock \& Dodds model for the non-linear power spectrum.
We assume $\Omega_\Lambda+\Omega_{\rm m}=1$, $\Gamma=\Omega_{\rm m}\,h$, $n=1$, $\Omega_{\rm m} \in [0,1]$ and marginalise over the Hubble parameter and the redshift distribution as described in the text.
{\bf Large panel}: black contours: cosmic shear analysis using the two-point shear-correlation function, light grey contours: aperture mass dispersion (shown are the $1\sigma$, $2\sigma$ and $3\sigma$ contours).
{\bf Upper right panel}: solid line: PDF of $\Omega_{\rm m}$ for the marginalisation over $\sigma_8$, dashed line: PDF of $\Omega_{\rm m}$ for $\sigma_8=0.8$.
{\bf Lower right panel}: solid line: PDF of $\sigma_8$ for the marginalisation over $\Omega_{\rm m}$, dashed line: PDF of $\sigma_8$ for $\Omega_{\rm m}=0.3$.
}
\label{fig:banane1}
\end{figure}

\begin{figure}
\centering
\includegraphics[scale=0.37,angle=-90]{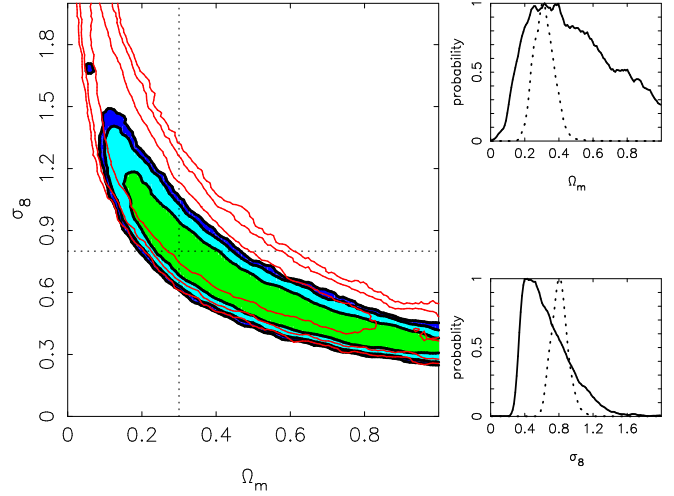}
\caption{
Model {\bf A}: same as in Fig. \ref{fig:banane1} but now for the aperture mass dispersion (black contours).
Light grey contours: comparison to the two-point shear-correlation function.
}
\label{fig:banane2}
\end{figure}

\paragraph{Estimated parameters. }
For a flat universe ($\Lambda$CDM model {\bf A}) both the aperture mass dispersion and the two-point shear correlation function is used to constrain cosmological parameters.
For $\langle M_{\rm ap}^2\rangle$ and $\xi_\pm$ we ran four independent MCMCs resulting in 15000 points (excluding 1500 points of the burn-in phase) and 29400 points (excluding 1000 points of the burn-in phase), respectively.
As a result we obtain the joint constraints on $\Omega_{\rm m}$ and $\sigma_8$, see Fig. \ref{fig:banane1} and Fig. \ref{fig:banane2}.
The confidence contours of $\sigma_8$ and $\Omega_{\rm m}$ reveal the typical ``banana''-like shape reflecting the strong degeneracy between these two parameters.
Due to the degeneracy both parameters are poorly constrained without further priors, see table \ref{tab:parameter} and solid lines in the small panels of Fig. \ref{fig:banane1} and Fig. \ref{fig:banane2}.

Assuming, as a strong prior, either $\Omega_{\rm m}=0.3$ or $\sigma_8=0.8$, we obtain stronger constrains on $\sigma_8$ and $\Omega_{\rm m}$, respectively, see table \ref{tab:parameter} and dashed lines in the small panels of Fig. \ref{fig:banane1} and Fig. \ref{fig:banane2}.  
\begin{table}
\caption{
Joint constraints on $\Omega_{\rm m}$ and $\sigma_8$ from our cosmic shear analysis using the two-point shear-correlation function and the aperture mass dispersion.
We assume $\Omega_\Lambda+\Omega_{\rm m}=1$, $\Gamma=\Omega_{\rm m}\,h$, $n=1$, $\Omega_{\rm m} \in [0,1]$ and marginalise over the Hubble parameter and the redshift distribution as described in the text.
Stated below are medians and $1\,\sigma$ errors.
}
\label{tab:parameter}
\begin{center}
    \begin{tabular}{c|c|c|c|c}
      \hline \hline
                                   &  $\Omega_{\rm m}$        & $\sigma_8$          & $\!\sigma_8\!(\Omega_{\rm m}\!=\!0.3)\!$ & $\!\Omega_{\rm m}\!(\sigma_8\!=\!0.8)\!$ \\
           \hline 
 \rule[-2mm]{0mm}{6mm} $\xi_\pm$   &  $0.44^{+0.45}_{-0.20}$ & $0.88^{+0.55}_{-0.42}$ &  $0.93 \pm 0.14$ & $0.37 \pm 0.08$\\
 \rule[-2mm]{0mm}{5mm}     $\langle M_{\rm ap}^2\rangle$ & $0.46^{+0.30}_{-0.22}$ & $0.61^{+0.32}_{-0.20}$ &  $0.80 \pm 0.10$ & $0.31 \pm 0.07$\\
      \hline
    \end{tabular}
\end{center}
\end{table}
%
%
%
%
%

\begin{figure}
\centering
\includegraphics[scale=0.38,angle=-90]{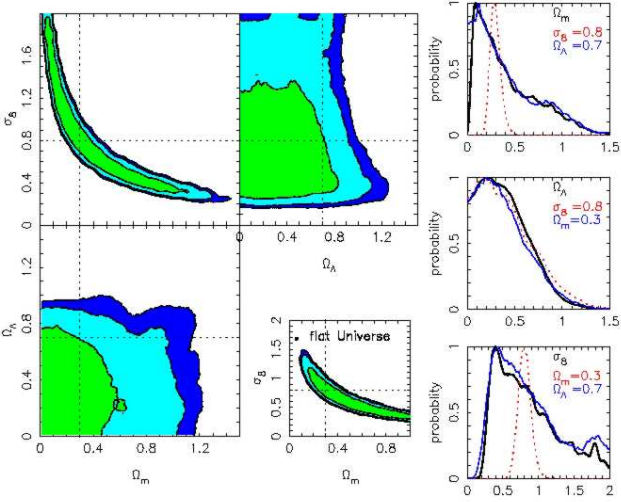}
\caption{
Model {\bf B}: {\bf bigger panels on the left}: joint constraints on $\Omega_{\rm m}$, $\sigma_8$ and $\Omega_\Lambda$ using the aperture mass dispersion for the $\Lambda$-universe: $\Omega_\Lambda$, $\Omega_{\rm m}\in[0,1.5]$.
{\bf Upper right panel:} black line: PDF of $\Omega_{\rm m}$ when marginalising over $\sigma_8$ and $\Omega_\Lambda$. 
Grey line: PDF of $\Omega_{\rm m}$ for $\Omega_\Lambda=0.7$.
Dotted line: PDF of $\Omega_{\rm m}$ for $\sigma_8=0.8$.
{\bf Centre right panel:} black line: PDF of $\Omega_\Lambda$ when marginalising over $\sigma_8$ and $\Omega_{\rm m}$. 
Grey line: PDF of $\Omega_\Lambda$ for $\Omega_{\rm m}=0.3$.
Dotted line: PDF of $\Omega_\Lambda$ for $\sigma_8=0.8$.
{\bf Lower right panel:} black line: PDF of $\sigma_8$ when marginalising over $\Omega_{\rm m}$ and $\Omega_\Lambda$. 
Grey line: PDF of $\sigma_8$ for $\Omega_\Lambda=0.7$.
Dotted line: PDF of $\sigma_8$ for $\Omega_{\rm m}=0.3$.
{\bf Small panel in the lower middle}: comparison to model {\bf A} (flat universe): joint constraints on $\Omega_{\rm m}$, $\sigma_8$.
}
\label{fig:banane3}
\end{figure}

For the $\Lambda$CDM model {\bf B} the aperture mass dispersion is utilised to constrain cosmological parameters.
We ran eight independent MCMCs resulting in 39000 points (excluding 1000 points of the burn-in phase).
The result is given in table \ref{tab:parameter2} and displayed in Fig. \ref{fig:banane3}.
Again, the confidence contours of $\sigma_8$ and $\Omega_{\rm m}$ reveal the typical ``banana''-like shape.
The contours of model {\bf B} are almost identical to model {\bf A}($M_{\rm ap}$) for $\Omega_{\rm m}>0.3$ and differ strongly for $\Omega_{\rm m}<0.3$ (the contours do not ``close'' until $\sigma_8\approx 1.9$).

Assuming as a strong prior either $\Omega_{\rm m}=0.3$ or $\sigma_8=0.8$ we obtain stronger constrains on $\sigma_8$ and $\Omega_{\rm m}$, respectively which is not the case for $\Omega_\Lambda$, see table \ref{tab:parameter2} and grey and light grey lines in the small panels of Fig. \ref{fig:banane3}.
Assuming as a strong prior $\Omega_\Lambda=0.7$ the constraints do not improve.
Hence the dependence on $\Omega_\Lambda$ is relatively weak, imposing only weak constraints.

However, a joint analysis of cosmic shear and supernovae measurements without assuming a flat universe (without priors from WMAP) could provide a tight constraint on $\Omega_\Lambda$.
Our measurements exclude a $\Omega_\Lambda$ larger than unity with $3\,\sigma$ and the supernovae measurement from \citet{kaa03} require a dark energy density ($\Omega_\Lambda>0$) with more than $3\,\sigma$.
\begin{table}
\caption{
Joint constraints on $\Omega_{\rm m}$, $\sigma_8$ and $\Omega_\Lambda$ from our cosmic shear analysis using the aperture mass dispersion.
We assume $\Omega_\Lambda$, $\Omega_{\rm m}\in[0,1.5]$, $\Gamma=\Omega_{\rm m}\,h$, $n=1$ and marginalise over the Hubble parameter and the redshift distribution as described in the text.
Stated below are medians and $1\,\sigma$ errors.
}
\label{tab:parameter2}
\begin{center}
    \begin{tabular}{c|c|c|c}
      \hline \hline
       add. prior                                   &  $\Omega_{\rm m}$      & $\sigma_8$            & $\Omega_\Lambda$ \\
      \hline
        \rule[-2mm]{0mm}{6mm} none                  &  $0.31_{-0.21}^{+0.45}$ & $0.78_{-0.36}^{+0.62}$ & $0.36_{-0.23}^{+0.30}$ \\ 
        \rule[-2mm]{0mm}{5mm} $\Omega_{\rm m}=0.3$   &          --           & $0.79_{-0.08}^{+0.09}$  & $0.33_{-0.22}^{+0.33}$ \\
        \rule[-2mm]{0mm}{5mm} $\sigma_8=0.8$        &  $0.29_{-0.05}^{+0.06}$ &          --           & $0.37_{-0.26}^{+0.35}$ \\
        \rule[-2mm]{0mm}{5mm} $\Omega_{\Lambda}=0.7$ &  $0.31_{-0.22}^{+0.49}$ & $0.76_{-0.36}^{+0.67}$ & --\\
      \hline
    \end{tabular}
\end{center}
\end{table}
\paragraph{Estimating parameters with $\langle M_{\rm ap}^2\rangle$ or $\xi_\pm$ ? }
The difference in the constraints on $\Omega_{\rm m}$ and $\sigma_8$, stemming from the applied aperture mass dispersion $\langle M_{\rm ap}^2\rangle$ and two-point shear correlation function $\xi_\pm$, can arise from the fact that both methods probe different parts of the mass power spectrum.  
Another reason for the difference arises from the fact that we fit $\langle M_{\rm ap}^2\rangle$ in the interval $\theta_0 \in [4^\prime,20^\prime]$ and $\xi_\pm$ in the interval $\theta_0 \in [0\myarcmin 8,33^\prime]$.

The aperture mass dispersion unambiguously separates E- and B-modes in the considered range, $\theta_0 \in [4^\prime,20^\prime]$ hence this method guarantees the revelation of potential systematics in contrast to $\xi_\pm$.
As our data is B-mode free for \mbox{$\theta_0>4\,{\rm arcmin}$}, the influence of systematics on the calculated E-mode signal is negligible within the interval $\theta_0 \in [4^\prime,20^\prime]$.
We conclude that our cosmological parameter estimation using $\langle M_{\rm ap}^2\rangle$ is therefore preferable to $\xi_\pm$.
%
%
%
\subsection{Systematic errors}
\label{sec:errors}
\citet{bvm97} showed that for a power law power spectrum the cosmological parameters $\sigma_8$ and $\Omega_{\rm m}$ are related to the top-hat shear variance $\langle \gamma^2\rangle$ (see Eq. \ref{eq:tophatshear}) and the mean source redshift $\bar z_{\rm s}$ as:
\be
\sigma_8^2 \, \Omega_{\rm m}^{1.7} \propto \langle \gamma^2\rangle \, \bar z_{\rm s}^{-1.7}.
\label{eq:estimation}
\ee
In the following we utilise this formula to obtain a rough systematic error estimate of $\sigma_8$.
\paragraph{Redshift distribution.  }
The redshift of source galaxies has a large impact on the estimation of cosmological parameters, see Eq. (\ref{eq:estimation}).
It is therefore crucial to know the redshift distribution accurately.
See \citet{htb06} and \citet{vwh06} for recent studies of photometric redshift errors and their influence on cosmic shear analyses.

Unlike most of the other cosmic shear surveys we estimate our redshift distribution from a sub-sample of the total lensing galaxy catalogue and not from an external redshift sample.
This has two advantages.
First, a redshift distribution of pure {\it lensing} galaxies is obtained.
Second, the sample variance can be estimated from the field-to-field variance of the redshift distributions obtained from single fields and can be taken into account in the parameter fit.

The sources of error of the redshift distribution are: 
a) the sample variance, as we only have redshift information available for $8\,\%$ of our lensing galaxies, and
b) the uncertain redshift information especially for those galaxies with \mbox{$R>23\,{\rm mag}$}.

To obtain a more reliable redshift distribution, about $7\,\%$ of the faint galaxies with highly uncertain redshifts are rejected using the aforementioned filter according to which the {\tt hyper-z} $68\,\%$-confidence interval should be smaller than $0.55\,(1+z)$ (note that the {\tt hyper-z}-uncertainties do not follow a Gaussian distribution).
With this filter the mean redshift is $8\,\%$ lower than the mean redshift of the unfiltered redshift distribution, since most of the rejected redshifts lie in the interval $[0.8,2.4]$, see Fig. \ref{fig:test_conf}.
However, as there are no spectroscopic or deep photometric redshift information obtained from bands which also comprise the infrared available it is impossible to judge if the rejected galaxies are really high-redshift galaxies or, for instance, dwarf galaxies at low redshift.
Thus it is not possible to say if we underestimate the mean redshift of our redshift distribution applying the aforementioned filter.
Comparing our measured mean redshift with that from \citet{blp06} obtained from galaxies in the same magnitude range we are confident that at least the mean redshift of our estimated redshift distribution is not strongly biased. 

To obtain a rough estimate of the systematic error in $\sigma_8$ due to the uncertain redshift distribution we simply assume the error of the mean redshift to be of the order of $8\,\%$.
This results approximately in an bias of $5\,\%$ for $\sigma_8$, see Eq. (\ref{eq:estimation}). 
\paragraph{Intrinsic alignment \& intrinsic shape-shear correlation.  }
The intrinsic alignment of close galaxy pairs introduces only a small contamination of the cosmic shear signal on small angular scales (\mbox{$\theta<1.5\,{\rm arcmin}$}) \citep{hwh06}.
The contamination could be removed totally by using redshift information \citep{kis03,heh03}.
As we do not have multi-colour information for most of our fields we cannot correct properly for this effect. 
However, the intrinsic alignment contamination is probably negligible as we average over a wide range in redshift and only probe angular scales with \mbox{$\theta > 4\,{\rm arcmin}$} (in the case of the $M_{\rm ap}$-statistics).

A further source of bias is the intrinsic shape-shear correlation predicted by \citet{his04}, measured by \citet{msk06} and analysed with numerical simulations by \citet{hwh06}.
\citet{hwh06} predicted that for a survey depth comparable to ours, the intrinsic shape-shear correlation produces a small negative B-mode and reduces the cosmic shear signal of the order of $15\,\%$ on scales below $20\,{\rm arcmin}$ resulting in an underestimation of $\sigma_8$ by about $7\,\%$.
\paragraph{Model.  }
Our estimation of cosmological parameters is based on the fitting formula proposed by \citet{ped96}.
\citet{hmv05} performed a cosmic shear analysis and obtained a $\sigma_8$-value using \citet{spj03} which is about $3.5\,\%$ smaller compared to the $\sigma_8$-value obtained using Peacock \& Dodds.
As our survey is comparable with \citet{hmv05} we assume that we overestimate $\sigma_8$ by about $4\,\%$ using Peacock \& Dodds (assuming the Smith et al.  model to be closer to reality).
\paragraph{PSF correction.  }
The presence of non-vanishing B-modes is a good indicator for systematics. 
However, \citet{hoe04} pointed out that the influence of systematics on the cosmic shear signal (E-mode) is not necessarily the same as on the B-modes.
This is probably the case in our analysis as indicated in Fig. \ref{fig:Sterne_ani}.
The PSF-anisotropy pattern of the WFI-fields have on average a larger impact on stellar E-modes than on stellar B-modes. 
In addition, we saw in Sect. \ref{sec:signal} that the B-mode signal is zero and does not suffer from an imperfect anisotropy correction but the E-mode signal probably does.
We concluded that the systematic error of the E-mode signal for angular scales larger than four arcminutes due to an imperfect anisotropy correction is in any case significantly lower compared to the statistical errors.

The results in STEP 2 indicate that on average we overestimate the shear by $3\,\%$ resulting in an overestimation of the E-mode signal.
In contrast, our cosmic shear analysis of synthetic {\tt skymaker}-images in Sect. \ref{sect:simulations} suggests that we underestimate the cosmic shear signal by a few percent.
Both, the probably small bias of the anisotropy correction and the bias in the shear estimation are simply accounted for by increasing the systematic error of $\sigma_8$ by $\pm 5\,\%$.

That the average B-mode signal is consistent with zero on scales larger than four arcminutes does not imply that single fields are B-mode free.
It is possible that B-modes resulting from different systematic errors only average out to zero.
If this is the case, the average E-mode value would be the same as without any systematic errors assuming that systematic errors bias the E-modes in the same way as B-modes.   
However, the full potential of the cosmic shear analysis is not achieved in this case as the errors would be larger than without any systematics.
In our measurements, the error bars of the average B-mode signal for scales larger than 6 arcminutes are smaller than those of the E-mode signal.
This is expected since B-modes are not affected by cosmic variance.
Hence, single fields are not dominated by systematic effects and we do not include this possible increase of the systematic errors into the final error estimation.
\paragraph{Field selection.  }
Our data set comprises one field with a massive cluster (A901) which was known before selecting the fields.
In addition, a field like A901 is unlikely to be part of a randomly chosen field sample compared to ours.
This can be construed as a prior knowledge which introduces a bias on the cosmic shear signal.
Replacing the E-mode signal of this field by the average E-mode signal, the final signal decreases by roughly $3.5\,\%$. 
This results in an potential overestimation of $\sigma_8$ by approximately $2\,\%$.
\paragraph{Error summary.  }
Taking all sources of systematic errors into account we see that we already entered a new phase of cosmic shear surveys.
That is, the systematic errors are of the same order of magnitude as statistical errors.
We obtain for the mass power spectrum normalisation assuming $\Omega_{\rm m}=0.3$ and including the statistical error and the rather conservative estimates of systematic errors: \mbox{$\sigma_8=0.80 (\pm 0.10 \pm 0.04 \pm 0.04 +0.06 -0.03 -0.01$)} ($1\,\sigma$ statistical error, redshift bias, PSF correction bias, intrinsic shape-shear correlation bias, model bias, field selection bias).
All given estimates of the systematic errors are maximum systematic errors. 
%
%
%
%
\section{Summary and outlook}
\label{sect:conclusion}
We have performed a cosmic shear analysis based on $15\,{\rm deg}^2$ of deep high quality $R$-band imaging data from the Garching-Bonn Deep Survey.
Our cosmic shear measurement is B-mode free within the statistical errors, hence there are no significant systematic errors resulting from the data treatment (PSF correction, galaxy selection) left in the data.
This encourages us to use the GaBoDS data for further cosmological analyses, such as the galaxy bias studies carried out by Simon et al. (in preparation). 

The measured redshift distribution is obtained from {\it lensing} galaxies from $2\,{\rm deg}^2$ of {\it UBVRI}-band data of the Deep Public Survey, a sub-sample of GaBoDS and not from an external redshift sample. 

\begin{figure}
\begin{center}
  \resizebox{\hsize}{!}{\includegraphics[width=\textwidth,clip]{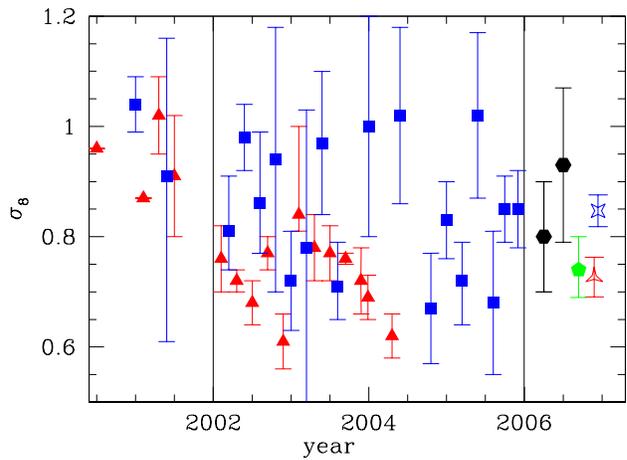}}
  \caption{
      Recent determinations of $\sigma_8(\Omega_{\rm m}=0.3)$ using galaxy clusters (triangles) and cosmic shear (squares) in comparison to our results ($M_{\rm ap}$: solid hexagon with bold error bars, $\xi_\pm$: hexagon with light error bars) and the WMAP three year result (pentagon).
      The open triangle and open star on the right are the average of all $\sigma_8$ determinations between 2002 and 2006 (indicated by the vertical lines) using clusters and cosmic shear, respectively (error bars of single measurements are not taken into account).
      The WMAP result of $\sigma_8$ would be larger if $\Omega_{\rm m}$ is fixed to $0.3$.
      Within a year the measurement points are not in chronological order.
      Values taken from table \ref{tab:sigma8vergleich}.
      Points from right to the left in the diagram are associated with values from top down in the table.
      \label{fig:sigma8vergleich}
}
\end{center}
\end{figure}

The measured redshift distribution in combination with the cosmic shear signal is used to perform an unbiased estimate of cosmological parameters where, in contrast to many other cosmic shear analyses, we have estimated the covariance matrix {\it directly from the data} without any further assumptions and have not computed it analytically. 
Assuming a flat $\Lambda$CDM universe with negligible baryon content we have derived the mass power spectrum normalisation and the matter density while marginalising over the uncertainties in the Hubble parameter and the source redshift distribution.
As a result we obtain \mbox{$\sigma_8=0.61^{+0.31}_{-0.20}$} and \mbox{$\Omega_{\rm m}=0.46^{+0.30}_{-0.22}$} from the $M_{\rm ap}$-statistics using the Peacock \& Dodds model of the non-linear mass power spectrum.
For a fixed matter density of \mbox{$\Omega_{\rm m}=0.3$}, we obtain \mbox{$\sigma_8=0.80 \pm 0.10$} ($1\,\sigma$ statistical error).
Within the error bars this is consistent with recent cosmic shear measurements, galaxy clusters and the WMAP three year result, see Fig. \ref{fig:sigma8vergleich} and table \ref{tab:sigma8vergleich}.

We discussed various systematic errors and roughly estimated their magnitude.
With respect to the magnitude of systematic errors, the most uncertain sources are the redshift distribution and the intrinsic shape-shear correlation. 
Although the intrinsic shape-shear correlation has been analysed with $N$-body simulations using simple galaxy models, the impact on deep cosmic shear measurements is still unclear.
With accurate photometric redshifts at hand, measurements of cross-correlation tomography could be carried out to provide a useful diagnostic tool \citep{his04}.
To reach the full potential of weak gravitational lensing measurements and high accuracy in the determination of cosmological parameters, it is therefore essential to have precise redshift estimates to obtain: a) unbiased redshift distributions of lensing galaxies and b) estimates of the intrinsic shape-shear correlation.

Our cosmic shear result and the ongoing improvements in redshift and shear estimates (e.g. STEP) are quite encouraging for the upcoming wide-field multi-colour surveys as the future KIlo Degree Survey (KIDS), starting beginning 2007.
The KIDS aims to image a contiguous area of $1500\,{\rm deg}^2$ in five colours.
In combination with UKIDS the data will yield photometric redshifts accurate to $\Delta z/(1+z)=0.03$ for typical  $r^\prime=23.5$ galaxies, rising to $10\,\%$ uncertainty at $r^\prime=25$.
In the primary lensing colour the limiting magnitude is predicted to be $r^\prime_{\rm Vega}=24.3$ ($10\,\sigma$ sky level measured in a circular aperture of $2^{\prime \prime}$ radius) and the median seeing is predicted to be $0\myarcsec 6$.
The depth of KIDS will yield about $10^8$ galaxies to a median redshift of $z\approx 0.8$.

With the KIDS data set at hand we will be able to measure the angular power spectrum on large scales.
The statistical errors of cosmological parameter estimations will be reduced at least by a factor of ten, not only because the number of background galaxies is larger but also the number of galaxy pairs will increase dramatically, in particular for large angular scales.
With such a large data set one can start to probe the equation of state of dark energy.
\begin{table}
  \caption{
    Recent determinations of $\sigma_8$ using clusters, cosmic shear and WMAP. 
    The WMAP result of $\sigma_8$ would be larger if $\Omega_{\rm m}$ is fixed to $0.3$.
    The result from \citet{crh04} is obtained from a radio survey assuming that the median redshift is $z_{\rm m}=2.2$.
  }
  \label{tab:sigma8vergleich}
  \begin{center}
    \begin{tabular}{l|c|l}
      \hline\hline
      Reference & $\sigma_8$ & Method \\
      \hline
      our measurements & $0.80 \pm 0.10$ & WL: $\langle M_{\rm ap}^2\rangle$ \\
                       & $0.93 \pm 0.14$ & WL: $\xi_\pm$ \\
      \citet{smv06} & $0.85 \pm 0.06$  & WL: $\xi^{\rm E}$  \\
      \citet{hmv05} & $0.85 \pm 0.05$  & WL: $\xi^{\rm E}$\\
      \citet{mrb05} & $1.02 \pm 0.15$ & WL: $\xi_\pm$ \\
      \citet{jjb05} & $0.72^{+0.08}_{-0.07}$ & WL: \\
      \citet{hbb05} & $0.68 \pm 0.13$  & WL: $\xi_\pm$\\
      \citet{vmh05} & $0.83 \pm 0.07$ & WL: $\xi^{\rm E}$\\
      \citet{hbh04} & $0.67 \pm 0.10$ & WL: $$\\
      \citet{rrc04} & $1.02 \pm 0.16$ & WL: $\langle \gamma^2\rangle$ \\
      \citet{crh04} & $1.0 \pm 0.2$   & WL: $\langle M_{\rm ap}^2\rangle$\\
      \citet{bmr03} & $0.97 \pm 0.13$ & WL: \\
      \citet{btb03} & $0.72 \pm 0.09$ & WL: \\
      \citet{hms03} & $0.78^{+0.55}_{-0.25}$ & WL: $\langle M_{\rm ap}^2\rangle$ \\
      \citet{jbf03} & $0.71^{+0.06}_{-0.08}$ & WL: \\
      \citet{hyg02} & $0.86^{+0.09}_{-0.13}$ & WL: $\langle M_{\rm ap}^2\rangle$  \\
      \citet{hyg02b}& $0.81^{+0.07}_{-0.10}$ & WL: $\langle \gamma^2\rangle$ \\
      \citet{vmp02} & $0.98 \pm 0.06$ & WL: $\langle M_{\rm ap}^2\rangle$ \\ 
      \citet{rrg02} & $0.94 \pm 0.24$ & WL:  \\
      \citet{rrg01} & $0.91 ^{+0.25}_{-0.30}$ & WL: \\
      \citet{mvm01} & $1.04 \pm 0.05$ & WL:   \\
      \hline
      \citet{hen04} & $0.62 \pm 0.04$ & X-ray clusters \\
      \citet{asf03} & $0.70 \pm 0.04$ & X-ray clusters \\
      \citet{bdb03} & $0.72 \pm 0.06$ & X-ray clusters  \\
      \citet{sbc03} & $0.76 \pm 0.01$ & X-ray clusters \\
      \citet{pbd03} & $0.77 \pm 0.05$ & X-ray clusters \\
      \citet{pbd03} & $0.78 \pm 0.06$ & X-ray clusters \\
      \citet{vkl03} & $0.84^{+0.16}_{-0.03}$ & X-ray clusters  \\
      \citet{vnl02} & $0.61 \pm 0.05$ & X-ray clusters \\
      \citet{reb02} & $0.68 \pm 0.04$ & X-ray clusters  \\
      \citet{rbn02} & $0.72 \pm 0.02$ & X-ray clusters  \\
      \citet{sel02} & $0.76 \pm 0.06$ & X-ray clusters  \\
      \citet{wu01}  & $0.91 \pm 0.11$ & X-ray clusters \\
      \citet{pbd01} & $1.02 \pm 0.07$ & X-ray clusters  \\
      \citet{oua01} & $0.91$ & X-ray clusters  \\
      \citet{bsb00} & $0.96$ & X-ray clusters  \\
      \hline
      \citet{sbd06} & $0.74^{+0.05}_{-0.06}$ & WMAP \\
    \end{tabular}
  \end{center}
\end{table}
\begin{acknowledgements} 
We thank Daniel Hudson for his careful reading of the manuscript and Ludovic Van Waerbeke for his useful comments.
Furthermore, we thank Takashi Hamana for very kindly allowing us to use his simulations.
This work was supported 
by the German Ministry for Science and Education (BMBF) through DESY
under the project 05AV5PDA/3,
and by the Deutsche Forschungsgemeinschaft under the projects
SCHN 342/6--1, ER 327/2--1.
Tim Schrabback acknowledges financial support by the Studienstiftung des deutschen Volkes.
In addition, we acknowlege the support given by ASTROVIRTEL, a Project funded by the European
Commission under FP5 Contract No. HPRI-CT-1999-00081.
\end{acknowledgements}
%
%


\begin{thebibliography}{91}
\expandafter\ifx\csname natexlab\endcsname\relax\def\natexlab#1{#1}\fi

\bibitem[{Allen {et~al.}(2003)Allen, Schmidt, Fabian, \& Ebeling}]{asf03}
Allen, S.~W., Schmidt, R.~W., Fabian, A.~C., \& Ebeling, H. 2003, MNRAS, 342,
  287

\bibitem[{Bacon {et~al.}(2003)Bacon, Massey, Refregier, \& Ellis}]{bmr03}
Bacon, D., Massey, R.~J., Refregier, A., \& Ellis, R. 2003, MNRAS, 344, 673

\bibitem[{Bacon {et~al.}(2000)Bacon, Refregier, \& Ellis}]{bre00}
Bacon, D., Refregier, A., \& Ellis, R. 2000, MNRAS, 318, 625

\bibitem[{Bahcall {et~al.}(2003)Bahcall, Dong, Bode, {et~al.}}]{bdb03}
Bahcall, N., Dong, F., Bode, P., {et~al.} 2003, ApJ, 585, 182

\bibitem[{Bardeen {et~al.}(1986)Bardeen, Bond, Kaiser, \& Szalay}]{bar86}
Bardeen, J.~M., Bond, J.~R., Kaiser, N., \& Szalay, A.~S. 1986, ApJ, 304, 15

\bibitem[{Bartelmann \& Schneider(2001)}]{bas01}
Bartelmann, M. \& Schneider, P. 2001, Phys. Rep., 340, 291

\bibitem[{{Bernardeau} {et~al.}(1997){Bernardeau}, {van Waerbeke}, \&
  {Mellier}}]{bvm97}
{Bernardeau}, F., {van Waerbeke}, L., \& {Mellier}, Y. 1997, A\&A, 322, 1

\bibitem[{Bertin \& Arnouts(1996)}]{bea96}
Bertin, E. \& Arnouts, S. 1996, A\&AS, 117, 393

\bibitem[{Blanchard {et~al.}(2000)Blanchard, Sadat, Bartlett, \&
  Le~Dour}]{bsb00}
Blanchard, A., Sadat, R., Bartlett, J.~G., \& Le~Dour, M. 2000, A\&A, 362, 809

\bibitem[{Bolzonella {et~al.}(2000)Bolzonella, Miralles, \& Pell\'o}]{bmp00}
Bolzonella, M., Miralles, J.~M., \& Pell\'o, R. 2000, A\&A, 363, 476

\bibitem[{Brodwin {et~al.}(2006)Brodwin, Lilly, Porciani, {et~al.}}]{blp06}
Brodwin, M., Lilly, S.~J., Porciani, {et~al.} 2006, ApJS, 162, 20

\bibitem[{Brown {et~al.}(2003)Brown, Taylor, Bacon, {et~al.}}]{btb03}
Brown, M.~L., Taylor, A.~N., Bacon, D.~J., {et~al.} 2003, MNRAS, 341, 100

\bibitem[{Bruzual~A. \& Charlot(1993)}]{brc93}
Bruzual~A., G. \& Charlot, S. 1993, AJ, 402, 538

\bibitem[{{Chang} {et~al.}(2004){Chang}, {Refregier}, \& {Helfand}}]{crh04}
{Chang}, T.-C., {Refregier}, A., \& {Helfand}, D.~J. 2004, ApJ, 617, 794

\bibitem[{{Clowe} {et~al.}(2006){Clowe}, {Schneider}, {Arag{\'o}n-Salamanca},
  {Bremer}, {de Lucia}, {Halliday}, {Jablonka}, {Milvang-Jensen}, {Pell{\'o}},
  {Poggianti}, {Rudnick}, {Saglia}, {Simard}, {White}, \& {Zaritsky}}]{csa06}
{Clowe}, D., {Schneider}, P., {Arag{\'o}n-Salamanca}, A., {et~al.} 2006, A\&A,
  451, 395

\bibitem[{Coleman {et~al.}(1980)Coleman, Wu, \& Weedman}]{cww80}
Coleman, G.~D., Wu, C.-C., \& Weedman, D.~W. 1980, ApJS, 239, 898

\bibitem[{Crittenden {et~al.}(2001)Crittenden, Natarajan, Pen, \&
  Theuns}]{cnp01}
Crittenden, R.~G., Natarajan, P., Pen, U.-L., \& Theuns, T. 2001, ApJ, 559, 552

\bibitem[{Crittenden {et~al.}(2002)Crittenden, Natarajan, Pen, \&
  Theuns}]{cnp02}
Crittenden, R.~G., Natarajan, P., Pen, U.-L., \& Theuns, T. 2002, ApJ, 568, 20

\bibitem[{Erben {et~al.}(2005)Erben, Schirmer, Dietrich, Cordes, Haberzettl,
  Hetterscheidt, {et~al.}}]{esd05}
Erben, T., Schirmer, M., Dietrich, J.~P., {et~al.} 2005, AN, 326, 432

\bibitem[{Erben {et~al.}(2001)Erben, van Waerbeke, Bertin, Mellier, \&
  Schneider}]{ewb01}
Erben, T., van Waerbeke, L., Bertin, E., Mellier, Y., \& Schneider, P. 2001,
  A\&A, 366, 717

\bibitem[{Gelman \& Rubin(1992)}]{ger92}
Gelman, A. \& Rubin, D.~B. 1992, Statistical Science, 7, 457

\bibitem[{{Grazian} {et~al.}(2006){Grazian}, {Fontana}, {de Santis}, {Nonino},
  {Salimbeni}, {Giallongo}, {Cristiani}, {Gallozzi}, \& {Vanzella}}]{gfd06}
{Grazian}, A., {Fontana}, A., {de Santis}, C., {et~al.} 2006, A\&A, 449, 951

\bibitem[{Hamana {et~al.}(2003)Hamana, Miyazaki, Shimasaku, Furusawa, Doi,
  {et~al.}}]{hms03}
Hamana, T., Miyazaki, S., Shimasaku, K., {et~al.} 2003, ApJ, 597, 98

\bibitem[{Hamana {et~al.}(2004)Hamana, Takada, \& Yoshida}]{hty04}
Hamana, T., Takada, M., \& Yoshida, N. 2004, MNRAS, 350, 893

\bibitem[{Hastings(1970)}]{has70}
Hastings, W.~K. 1970, Biometrika, 57, 97

\bibitem[{Heavens {et~al.}(2000)Heavens, Refregier, \& Heymans}]{hrh00}
Heavens, A., Refregier, A., \& Heymans, C. 2000, MNRAS, 319, 649

\bibitem[{Henry(2004)}]{hen04}
Henry, J.~P. 2004, ApJ, 609, 603

\bibitem[{Hetterscheidt {et~al.}(2005)Hetterscheidt, Erben, Schneider, Maoli,
  Van~Waerbeke, \& Mellier}]{het05}
Hetterscheidt, M., Erben, T., Schneider, P., {et~al.} 2005, A\&A, 442, 43

\bibitem[{{Heymans} {et~al.}(2004){Heymans}, {Brown}, {Heavens},
  {Meisenheimer}, {Taylor}, \& {Wolf}}]{hbh04}
{Heymans}, C., {Brown}, M., {Heavens}, A., {et~al.} 2004, MNRAS, 347, 895

\bibitem[{{Heymans} {et~al.}(2005){Heymans}, {Brown}, {Barden}, {Caldwell},
  {Jahnke}, {Peng}, {Rix}, {Taylor}, {Beckwith}, {Bell}, {Borch},
  {H{\"a}u{\ss}ler}, {Jogee}, {McIntosh}, {Meisenheimer}, {S{\'a}nchez},
  {Somerville}, {Wisotzki}, \& {Wolf}}]{hbb05}
{Heymans}, C., {Brown}, M.~L., {Barden}, M., {et~al.} 2005, MNRAS, 361, 160

\bibitem[{Heymans \& Heavens(2003)}]{heh03}
Heymans, C. \& Heavens, A. 2003, MNRAS, 339, 711

\bibitem[{Heymans {et~al.}(2006{\natexlab{a}})Heymans, Van~Waerbeke, Bacon,
  {et~al.}}]{hvb06}
Heymans, C., Van~Waerbeke, L., Bacon, D., {et~al.} 2006{\natexlab{a}}, MNRAS,
  368, 1323

\bibitem[{Heymans {et~al.}(2006{\natexlab{b}})Heymans, White, Heavens, \&
  Van~Waerbeke}]{hwh06}
Heymans, C., White, M., Heavens, A., \& Van~Waerbeke, L. 2006{\natexlab{b}},
  preprint astro-ph/0604001

\bibitem[{Hildebrandt {et~al.}(2006)Hildebrandt, Erben, Dietrich, Cordes,
  Haberzettl, Hetterscheidt, {et~al.}}]{hed05}
Hildebrandt, H., Erben, T., Dietrich, J., {et~al.} 2006, A\&A, 452, 1121

\bibitem[{Hirata \& Seljak(2004)}]{his04}
Hirata, C.~M. \& Seljak, U. 2004, Phys.Rev.D, 063526

\bibitem[{Hoekstra(2004)}]{hoe04}
Hoekstra, H. 2004, MNRAS, 347, 1337

\bibitem[{Hoekstra {et~al.}(1998)Hoekstra, Franx, Kuijken, \& Squires}]{hfk98}
Hoekstra, H., Franx, M., Kuijken, K., \& Squires, G. 1998, ApJ, 504, 636

\bibitem[{Hoekstra {et~al.}(2005)Hoekstra, Mellier, Van~Waerbeke,
  {et~al.}}]{hmv05}
Hoekstra, H., Mellier, Y., Van~Waerbeke, L., {et~al.} 2005, astro-ph/0511089

\bibitem[{Hoekstra {et~al.}(2002)Hoekstra, Yee, \& Gladders}]{hyg02}
Hoekstra, H., Yee, H. K.~C., \& Gladders, M.~D. 2002, ApJ, 577, 595

\bibitem[{{Hoekstra} {et~al.}(2002b){Hoekstra}, {Yee}, {Gladders},
  {Barrientos}, {Hall}, \& {Infante}}]{hyg02b}
{Hoekstra}, H., {Yee}, H.~K.~C., {Gladders}, M.~D., {et~al.} 2002b, ApJ, 572,
  55

\bibitem[{{Huterer} {et~al.}(2006){Huterer}, {Takada}, {Bernstein}, \&
  {Jain}}]{htb06}
{Huterer}, D., {Takada}, M., {Bernstein}, G., \& {Jain}, B. 2006, MNRAS, 366,
  101

\bibitem[{{Jarvis} {et~al.}(2003){Jarvis}, {Bernstein}, {Fischer}, {Smith},
  {Jain}, {Tyson}, \& {Wittman}}]{jbf03}
{Jarvis}, M., {Bernstein}, G.~M., {Fischer}, P., {et~al.} 2003, AJ, 125, 1014

\bibitem[{Jarvis \& Jain(2004)}]{jaj04}
Jarvis, M. \& Jain, B. 2004, astro-ph/0412234

\bibitem[{{Jarvis} {et~al.}(2006){Jarvis}, {Jain}, {Bernstein}, \&
  {Dolney}}]{jjb05}
{Jarvis}, M., {Jain}, B., {Bernstein}, G., \& {Dolney}, D. 2006, ApJ, 644, 71

\bibitem[{Kaiser(1998)}]{kas98}
Kaiser, N. 1998, ApJ, 498, 26

\bibitem[{Kaiser {et~al.}(1995)Kaiser, Squires, \& Broadhurst}]{ksb95}
Kaiser, N., Squires, G., \& Broadhurst, T. 1995, ApJ, 449, 460

\bibitem[{Kaiser {et~al.}(2000)Kaiser, Wilson, \& Luppino}]{kwl00}
Kaiser, N., Wilson, G., \& Luppino, G. 2000, ApJL, submitted, preprint
  astro-ph/0003338

\bibitem[{Kilbinger {et~al.}(2006)Kilbinger, Schneider, \& Eifler}]{kse06}
Kilbinger, M., Schneider, P., \& Eifler, T. 2006, astro-ph/0604520

\bibitem[{{King} \& {Schneider}(2002)}]{kis02}
{King}, L. \& {Schneider}, P. 2002, A\&A, 396, 411

\bibitem[{{King} \& {Schneider}(2003)}]{kis03}
{King}, L. \& {Schneider}, P. 2003, A\&A, 398, 23

\bibitem[{{Knop} {et~al.}(2003){Knop}, {Aldering}, {Amanullah}, {Astier},
  {Blanc}, {Burns}, {Conley}, {Deustua}, {Doi}, {Ellis}, {Fabbro}, {Folatelli},
  {Fruchter}, {Garavini}, {Garmond}, {Garton}, {Gibbons}, {Goldhaber},
  {Goobar}, {Groom}, {Hardin}, {Hook}, {Howell}, {Kim}, {Lee}, {Lidman},
  {Mendez}, {Nobili}, {Nugent}, {Pain}, {Panagia}, {Pennypacker}, {Perlmutter},
  {Quimby}, {Raux}, {Regnault}, {Ruiz-Lapuente}, {Sainton}, {Schaefer},
  {Schahmaneche}, {Smith}, {Spadafora}, {Stanishev}, {Sullivan}, {Walton},
  {Wang}, {Wood-Vasey}, \& {Yasuda}}]{kaa03}
{Knop}, R.~A., {Aldering}, G., {Amanullah}, R., {et~al.} 2003, ApJ, 598, 102

\bibitem[{Le~F\`evre {et~al.}(2004)Le~F\`evre, Vettolani, Paltani, Tresse,
  Zamorani, Brun, {et~al.}}]{fvp04}
Le~F\`evre, O., Vettolani, G., Paltani, S., {et~al.} 2004, A\&A, 228, 1043

\bibitem[{{Ma} {et~al.}(2006){Ma}, {Hu}, \& {Huterer}}]{mhh06}
{Ma}, Z., {Hu}, W., \& {Huterer}, D. 2006, ApJ, 636, 21

\bibitem[{Mandelbaum {et~al.}(2006)Mandelbaum, Hirata, Ishak, Seljak, \&
  Brinkmann}]{msk06}
Mandelbaum, R., Hirata, C.~M., Ishak, M., Seljak, U., \& Brinkmann, J. 2006,
  MNRAS, 367, 611

\bibitem[{Maoli {et~al.}(2001)Maoli, van Waerbeke, Mellier, {et~al.}}]{mvm01}
Maoli, R., van Waerbeke, L., Mellier, Y., {et~al.} 2001, A\&A, 368, 766

\bibitem[{Massey {et~al.}(2005)Massey, Refregier, Bacon, Ellis, \&
  Brown}]{mrb05}
Massey, R., Refregier, A., Bacon, D., Ellis, R., \& Brown, M.~L. 2005, MNRAS,
  359, 1277

\bibitem[{Metropolis {et~al.}(1953)Metropolis, Rosenbluth, Rosenbluth, Teller,
  \& Teller}]{mrr53}
Metropolis, N., Rosenbluth, A.~W., Rosenbluth, M.~N., Teller, A., \& Teller, E.
  1953, Journal of Chemical Physics, 21, 1087

\bibitem[{Oukbir \& Arnaud(2001)}]{oua01}
Oukbir, J. \& Arnaud, M. 2001, MNRAS, 326, 453

\bibitem[{Peacock \& Dodds(1996)}]{ped96}
Peacock, J.~A. \& Dodds, S.~J. 1996, MNRAS, 280, 19

\bibitem[{{Pierfederici}(2001)}]{pie01}
{Pierfederici}, F. 2001, in Proc. SPIE Vol. 4477, p. 246-253, Astronomical Data
  Analysis, Jean-Luc Starck; Fionn D. Murtagh; Eds., ed. J.-L. {Starck} \&
  F.~D. {Murtagh}, 246--253

\bibitem[{Pierpaoli {et~al.}(2003)Pierpaoli, Borgani, Scott, \& White}]{pbd03}
Pierpaoli, E., Borgani, S., Scott, D., \& White, M. 2003, MNRAS, 342, 163

\bibitem[{Pierpaoli {et~al.}(2001)Pierpaoli, Scott, \& White}]{pbd01}
Pierpaoli, E., Scott, D., \& White, M. 2001, MNRAS, 325, 77

\bibitem[{Refregier {et~al.}(2002)Refregier, Rhodes, \& Groth}]{rrg02}
Refregier, A., Rhodes, J., \& Groth, E.~J. 2002, ApJ, 572, L131

\bibitem[{Reiprich \& B\"ohringer(2002)}]{reb02}
Reiprich, T.~H. \& B\"ohringer, H. 2002, ApJ, 567, 716

\bibitem[{Rhodes {et~al.}(2004)Rhodes, Refregier, Collins, Gardner, Groth, \&
  Hill}]{rrc04}
Rhodes, J., Refregier, A., Collins, N.~R., {et~al.} 2004, ApJ, 605, 29

\bibitem[{Rhodes {et~al.}(2001)Rhodes, Refregier, \& Groth}]{rrg01}
Rhodes, J., Refregier, A., \& Groth, E. 2001, ApJ, 552, L85

\bibitem[{Rosati {et~al.}(2002)Rosati, Borgani, \& Norman}]{rbn02}
Rosati, P., Borgani, S., \& Norman, C. 2002, ARA\&A, 40, 539

\bibitem[{Schirmer(2004)}]{sch04}
Schirmer, M. 2004, PhD thesis, Universit\"at Bonn, urn:nbn:de:hbz:5n-03263;
  URL:
  http://hss.ulb.uni-bonn.de/diss\_online/math\_nat\_fak/2004/schirmer\_mischa

\bibitem[{Schirmer {et~al.}(2003)Schirmer, Erben, Schneider, {et~al.}}]{ses03}
Schirmer, M., Erben, T., Schneider, P., {et~al.} 2003, A\&A, 407, 869

\bibitem[{Schneider(1996)}]{sch96_2}
Schneider, P. 1996, MNRAS, 283, 837

\bibitem[{{Schneider} \& {Kilbinger}(2006)}]{sck06}
{Schneider}, P. \& {Kilbinger}, M. 2006, astro-ph/0605084

\bibitem[{Schneider {et~al.}(2006)Schneider, Kochanek, \& Wambsganss}]{sch05}
Schneider, P., Kochanek, C., \& Wambsganss, J. 2006, Gravitational Lensing:
  Strong, Weak and Micro (Springer; Saas-Fee Advanced Courses)

\bibitem[{Schneider {et~al.}(1998)Schneider, Van~Waerbeke, Jain, \&
  Kruse}]{svj98}
Schneider, P., Van~Waerbeke, L., Jain, B., \& Kruse, G. 1998, MNRAS, 296, 873

\bibitem[{{Schneider} {et~al.}(2002){Schneider}, {van Waerbeke}, {Kilbinger},
  \& {Mellier}}]{svk02}
{Schneider}, P., {van Waerbeke}, L., {Kilbinger}, M., \& {Mellier}, Y. 2002,
  A\&A, 396, 1

\bibitem[{Schneider {et~al.}(2002)Schneider, Van~Waerbeke, \& Mellier}]{svm02}
Schneider, P., Van~Waerbeke, L., \& Mellier, Y. 2002, A\&A, 389, 729

\bibitem[{Schuecker {et~al.}(2003)Schuecker, B\"ohringer, Collins, \&
  Guzzo}]{sbc03}
Schuecker, P., B\"ohringer, H., Collins, C.~A., \& Guzzo, L. 2003, A\&A, 398,
  867

\bibitem[{Seljak(2002)}]{sel02}
Seljak, U. 2002, MNRAS, 337, 774

\bibitem[{{Semboloni} {et~al.}(2006){Semboloni}, {Mellier}, {van Waerbeke},
  {Hoekstra}, {Tereno}, {Benabed}, {Gwyn}, {Fu}, {Hudson}, {Maoli}, \&
  {Parker}}]{smv06}
{Semboloni}, E., {Mellier}, Y., {van Waerbeke}, L., {et~al.} 2006, A\&A, 452,
  51

\bibitem[{Smith {et~al.}(2003)Smith, Peacock, Jenkins, White, Frenk, Pearce,
  Thomas, Efstathiou, \& Couchman}]{spj03}
Smith, R.~E., Peacock, J.~A., Jenkins, A., {et~al.} 2003, MNRAS, 341, 1311

\bibitem[{Spergel {et~al.}(2006)Spergel, Bean, Dore', {et~al.}}]{sbd06}
Spergel, D.~N., Bean, R., Dore', O., {et~al.} 2006, preprint astro-ph/0603449

\bibitem[{Tereno {et~al.}(2005)Tereno, Dore, Van~Waerbeke, \& Mellier}]{tdv05}
Tereno, I., Dore, O., Van~Waerbeke, L., \& Mellier, Y. 2005, A\&A, 429, 383

\bibitem[{Van~Waerbeke {et~al.}(2000)Van~Waerbeke, Mellier, Erben,
  {et~al.}}]{vme00}
Van~Waerbeke, L., Mellier, Y., Erben, T., {et~al.} 2000, A\&A, 358, 30

\bibitem[{Van~Waerbeke {et~al.}(2005)Van~Waerbeke, Mellier, \&
  Hoekstra}]{vmh05}
Van~Waerbeke, L., Mellier, Y., \& Hoekstra, H. 2005, A\&A, 429, 75

\bibitem[{Van~Waerbeke {et~al.}(2002)Van~Waerbeke, Mellier, Pello, Pen,
  McCracken, \& Jain}]{vmp02}
Van~Waerbeke, L., Mellier, Y., Pello, R., {et~al.} 2002, A\&A, 393, 369

\bibitem[{Van~Waerbeke {et~al.}(2001)Van~Waerbeke, Mellier, Radovich, Bertin,
  Dantel-Fort, McCracken, \& others.}]{vmr01}
Van~Waerbeke, L., Mellier, Y., Radovich, M., {et~al.} 2001, A\&A, 374, 757

\bibitem[{Van~Waerbeke {et~al.}(2006)Van~Waerbeke, White, Hoekstra, \&
  Heymans}]{vwh06}
Van~Waerbeke, L., White, M., Hoekstra, H., \& Heymans, C. 2006,
  astro-ph/0603696

\bibitem[{Viana {et~al.}(2003)Viana, Kay, Liddle, Muanwong, \& Thomas}]{vkl03}
Viana, P. T.~P., Kay, S.~T., Liddle, A.~R., Muanwong, O., \& Thomas, P.~A.
  2003, MNRAS, 346, 319

\bibitem[{Viana {et~al.}(2002)Viana, Nichol, \& Liddle}]{vnl02}
Viana, P. T.~P., Nichol, R.~C., \& Liddle, A.~R. 2002, MNRAS, 336, 541

\bibitem[{{White} {et~al.}(2005){White}, {Clowe}, {Simard}, {Rudnick}, {de
  Lucia}, {Arag{\'o}n-Salamanca}, {Bender}, {Best}, {Bremer}, {Charlot},
  {Dalcanton}, {Dantel}, {Desai}, {Fort}, {Halliday}, {Jablonka}, {Kauffmann},
  {Mellier}, {Milvang-Jensen}, {Pell{\'o}}, {Poggianti}, {Poirier},
  {Rottgering}, {Saglia}, {Schneider}, \& {Zaritsky}}]{wcs06}
{White}, S.~D.~M., {Clowe}, D.~I., {Simard}, L., {et~al.} 2005, A\&A, 444, 365

\bibitem[{Wittman {et~al.}(2000)Wittman, Tyson, Kirkman, Dell'Antonio,
  Bernstein, {et~al.}}]{wtk00}
Wittman, D.~M., Tyson, J.~A., Kirkman, D., {et~al.} 2000, Nature, 405, 143

\bibitem[{{Wu}(2001)}]{wu01}
{Wu}, J.-H.~P. 2001, \mnras, 327, 629

\end{thebibliography}
%
\appendix
\section{Single fields}
\label{appendixa}
In this appendix we give the main characteristics of the single fields and display the relevant $M_{\rm ap}$-statistics and relative weighting factors of all fields.  
\begin{table*}
\caption{List of all fields of the GaBoDS survey. 
The quantity $N$ is the number of galaxies per field of the final catalogue in the magnitude range $R \in [21.5,24.5]$ and $2\,r_{\rm h}$ is twice the measured half light radius of stars in arcseconds.}
\label{SumTab}
\begin{center}
\begin{tabular}{l|r|r|r|r|r|l|l}
\hline \hline
name &RA (J2000) & DEC (J2000) & $t_{\rm obs}$[ksec] & N & $2\,r_{\rm h}[^{\prime\prime}]$ & data source & code \\
\hline
      A901      &149.07771 &$-$10.02734 &18.1 & 19513&0.75& COMBO-17 & COMBO\\
      AXAF      &53.13344  &$-$27.82255 &57.2 & 19639&0.90& COMBO-17 & COMBO\\
      FDF       &16.44541  &$-$25.85742 &11.8 & 19169&0.86& COMBO-17 & COMBO\\
      S11       &175.74860 &$-$1.73458  &21.5 & 18855&0.84& COMBO-17 & COMBO\\
      SGP       &11.49852  &$-$29.61047 &20.0 & 21465&0.85& COMBO-17 & COMBO\\
\hline                                    
      CAPO\_DF  &186.03787 &$-$13.10764 &13.0 & 16014&0.97& ESO archive & OWN\\
      NDF       &181.36237 &$-$7.65226  &21.8 & 14146 &0.93& ME IR group & OWN\\  
      SHARC2    &76.3333   &$-$28.81805 &11.4 & 19101&0.86& own obs. & OWN\\
      F17\_P1   &216.41916 &$-$34.69460 &10.0 & 10345&1.13& own obs. & OWN\\
      F17\_P3   &217.02611 &$-$34.69463 &10.0 & 14135 &0.77& own obs. & OWN\\
      A1347\_P1 &175.25702 &$-$25.51474 &13.5 & 10330 &0.76& own obs. & OWN\\
      A1347\_P2 &175.79254 &$-$25.50918 &7.5  & 13738 &0.91& own obs. & OWN\\
      A1347\_P3 &175.23976 &$-$25.00933 &7.0  & 15581 &0.89& own obs. & OWN\\
      A1347\_P4 &175.79459 &$-$24.99836 &8.0  & 13665 &1.07& own obs. & OWN\\
      F4\_P1    &321.65611 &$-$40.25193 &9.5  & 16135&0.93& own obs. & OWN\\
      F4\_P2    &321.71942 &$-$39.76761 &7.0  & 14103 &1.06& own obs. & OWN\\
      F4\_P3    &322.32389 &$-$39.72689 &10.0 & 16348 &0.87& own obs. & OWN\\
      F4\_P4    &322.32389 &$-$39.72689 &7.5  & 12943 &1.10& own obs. & OWN\\
\hline                                    
      Deep1a    &343.79506 &$-$40.19886 &7.2  & 16131&1.00& EIS & DEEP\\
      Deep1b    &343.058572&$-$40.22481& 3.9  & 9806 &1.30& EIS & DEEP\\
      Deep1c    &342.328125&$-$40.20702& 3.9  & 13354&1.03& EIS & DEEP\\
      Deep1e    &341.96679 &$-$39.52874 &9.0  & 13374&1.05& EIS & DEEP\\
      Deep2a    &54.372223 &$-$27.81551& 6.0  & 14785&1.00& EIS & DEEP\\
      Deep2b    &53.746626 &$-$27.80862& 5.1  & 10177&1.20& EIS & DEEP\\
      Deep2d    &52.506344 &$-$27.81774& 3.0  & 9745 &1.20& EIS & DEEP\\
      DEEP2e    &53.12291  &$-$27.30467 &7.5  & 16768&0.97& own obs. & DEEP\\
      DEEP2f    &53.66995  &$-$27.32400 &7.0  & 14494&1.10& own obs. & DEEP\\
      Deep3a    &171.24559 &$-$21.68289 &7.2  & 14130&0.91& EIS & DEEP\\
      Deep3b    &170.66159 &$-$21.70969 &9.3  & 14905&0.92& EIS & DEEP\\
      Deep3c    &170.01909 &$-$21.69960 &9.0  & 15606&0.90& EIS & DEEP\\
      Deep3d    &169.42875 &$-$25.85742 &9.3  & 14134&0.80& own obs. & DEEP\\
\hline                                    
      AM1         &58.81181  &$-$49.667762&7.5&14956 &0.99& ASTROVIRTEL & B8\\
      B8p0        &340.34886 &$-$9.59009  &7.2&13299 &0.87& ASTROVIRTEL & B8\\
      B8p1        &340.346051&$-$9.08957 & 4.5&13654 &0.87& ASTROVIRTEL & B8\\ 
      B8p2        &340.345309&$-$8.58963 & 5.4&13308 &0.87& ASTROVIRTEL & B8\\ 
      B8p3        &340.345149&$-$8.08951 & 5.4&13834 &0.89& ASTROVIRTEL & B8\\ 
      B8m1        &340.34884 &$-$10.08953& 4.5&12401 &0.90& ASTROVIRTEL & B8\\ 
      B8m2        &340.348548&$-$10.58954& 5.4&12864 &0.89& ASTROVIRTEL & B8\\ 
      B8m3        &340.346888&$-$11.08857& 5.4&14397 &0.99& ASTROVIRTEL & B8\\ 
      Comparison  &65.307669 &$-$36.28380& 5.3&14882 &1.02& ASTROVIRTEL & B8\\
      Pal3        &151.432117&$-$0.00344 & 4.2&12608 &1.07& ASTROVIRTEL & B8\\
\hline                                    
      C0400       &214.360609&$-$12.25356& 4.8&11857 &0.90& ASTROVIRTEL & C0\\ 
      C04m1       &214.727417&$-$12.75342& 4.0&10536 &0.94& ASTROVIRTEL & C0\\ 
      C04m2       &214.478576&$-$13.25319& 4.0&11292 &0.89& ASTROVIRTEL & C0\\ 
      C04m3       &215.318002&$-$13.75352& 4.0&13170 &0.85& ASTROVIRTEL & C0\\ 
      C04m4       &215.111423&$-$14.25337& 4.0&11872 &0.88& ASTROVIRTEL & C0\\ 
      C04p1       &214.726983&$-$11.75319& 4.0&11760 &0.92& ASTROVIRTEL & C0\\ 
      C04p2       &214.727266&$-$11.25366& 4.0&11093 &0.90& ASTROVIRTEL & C0\\ 
      C04p3       &215.098018&$-$10.75340& 4.0&13702 &0.85& ASTROVIRTEL & C0\\
\hline                                    
      CL1037-1243 &159.444072&$-$12.75499& 3.6& 13210&1.03& EDisCS & CL\\
      CL1040-1155 &160.139300&$-$11.96379& 3.6& 13451&0.99& EDisCS & CL\\
      CL1054-1146 &163.581888&$-$11.81304& 3.6& 12263&1.00& EDisCS & CL\\
      CL1054-1245 &163.647353&$-$12.79700& 3.6& 15420&0.90& EDisCS & CL\\
      CL1059-1253 &164.755650&$-$12.92051& 3.6& 12778&1.05& EDisCS & CL\\
      CL1119-1129 &169.784677&$-$11.52558& 3.6& 13635&0.99& EDisCS & CL\\
      CL1138-1133 &174.508878&$-$11.59953& 3.6& 16748&0.93& EDisCS & CL\\
      CL1202-1224 &180.645603&$-$12.44172& 3.6& 14935&0.97& EDisCS & CL\\
      Cl1216-1201 &184.170966&$-$12.06268& 3.6& 17472&0.75& EDisCS & CL\\  
      Cl1301-1139 &195.467853&$-$11.63099& 3.6& 13975&0.96& EDisCS & CL\\
      CL1353-1137 &208.306268&$-$11.59813& 3.6& 15281&0.90& EDisCS & CL\\
      CL1420-1236 &215.066773&$-$12.64986& 3.6& 11223&1.04& EDisCS & CL\\
\hline
\end{tabular}
\end{center}
\end{table*}
%
%
%
\begin{figure}
\centering
\resizebox{\hsize}{!}{\includegraphics[width=\textwidth,clip]{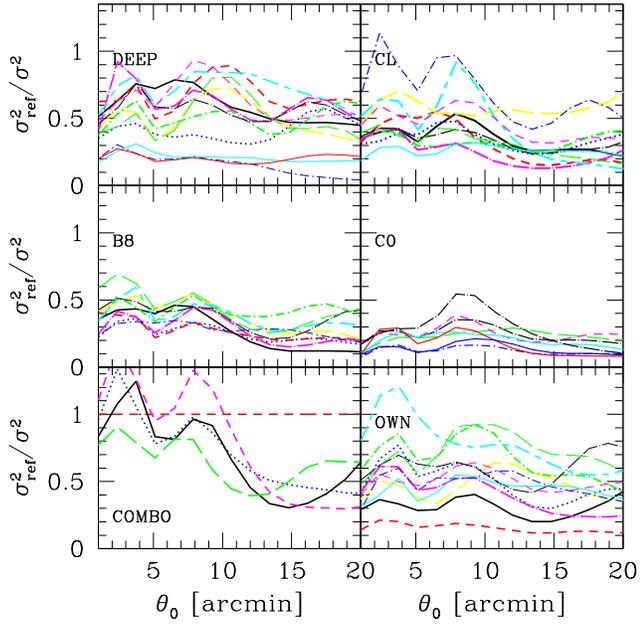}}
\caption{Relative statistical weights of the GaBoDS fields versus the aperture radius $\theta_0$ obtained via bootstrapping on galaxy basis.
The COMBO field AXAF is chosen as a reference field.
The type of each line is the same as in Fig. \ref{fig:singlefields}. 
}
\label{fig:weight}
\end{figure}
%
%
\begin{figure*}
\centering
\includegraphics[scale=0.38]{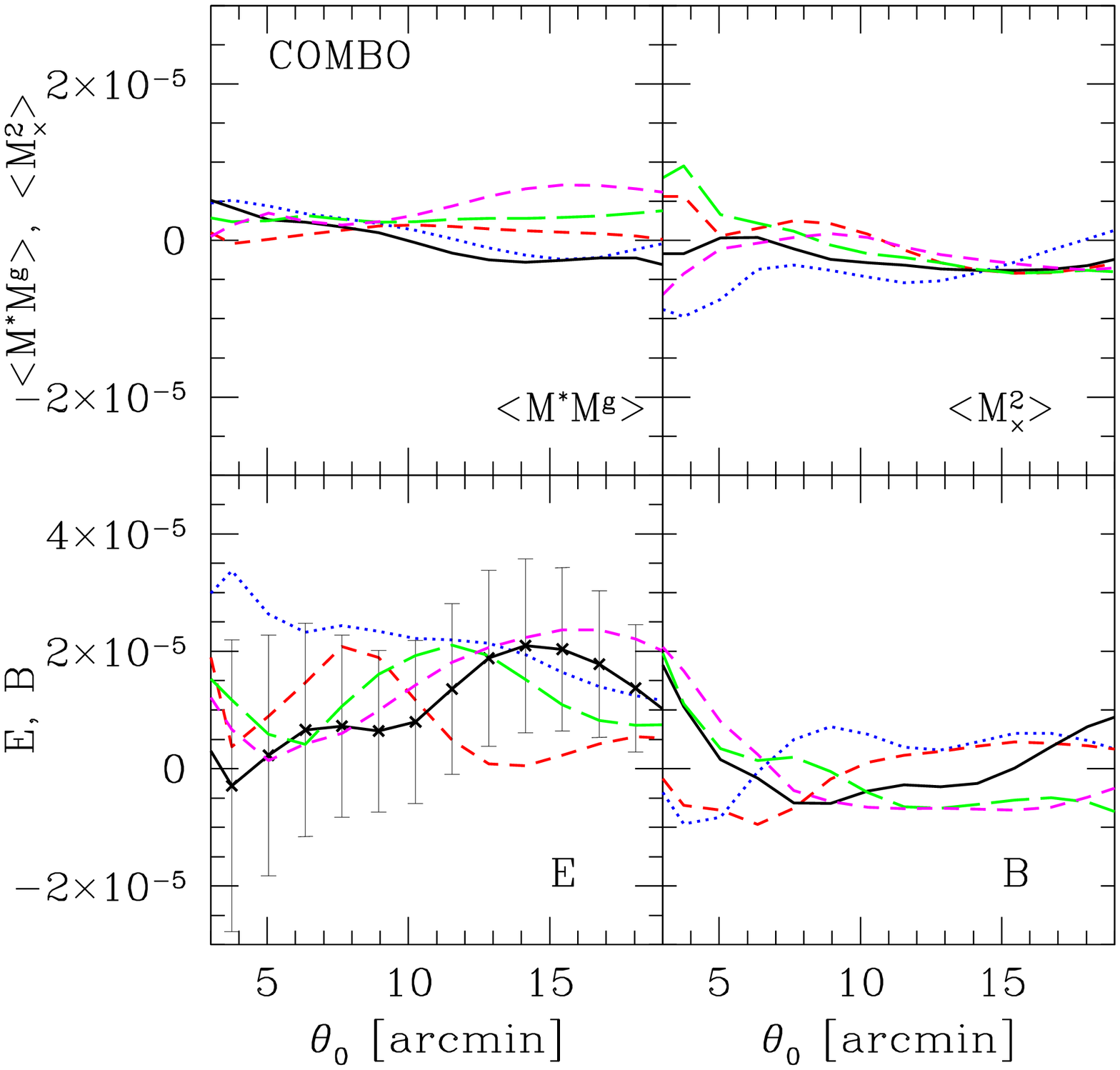}
\includegraphics[scale=0.38]{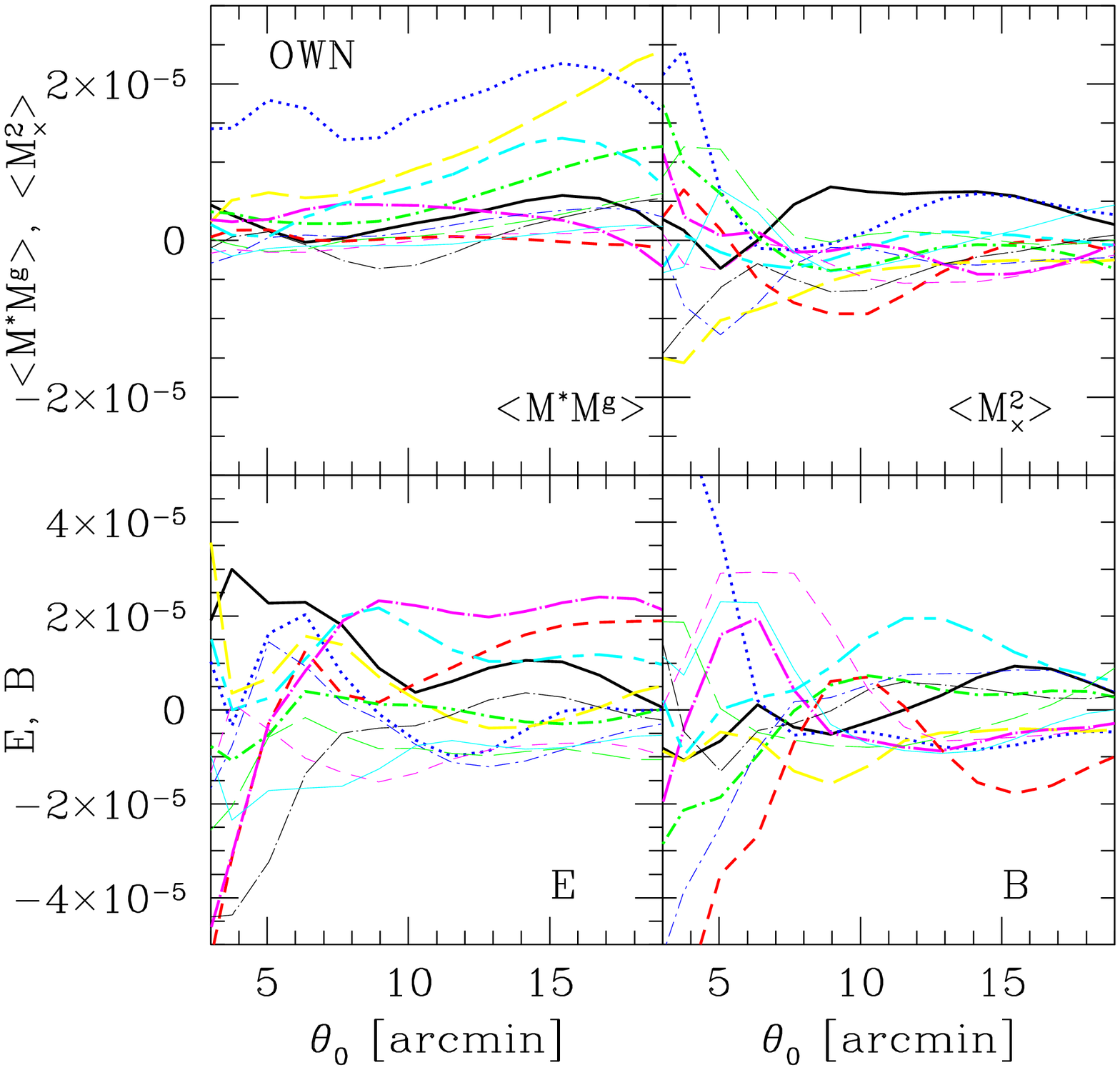}
\includegraphics[scale=0.38]{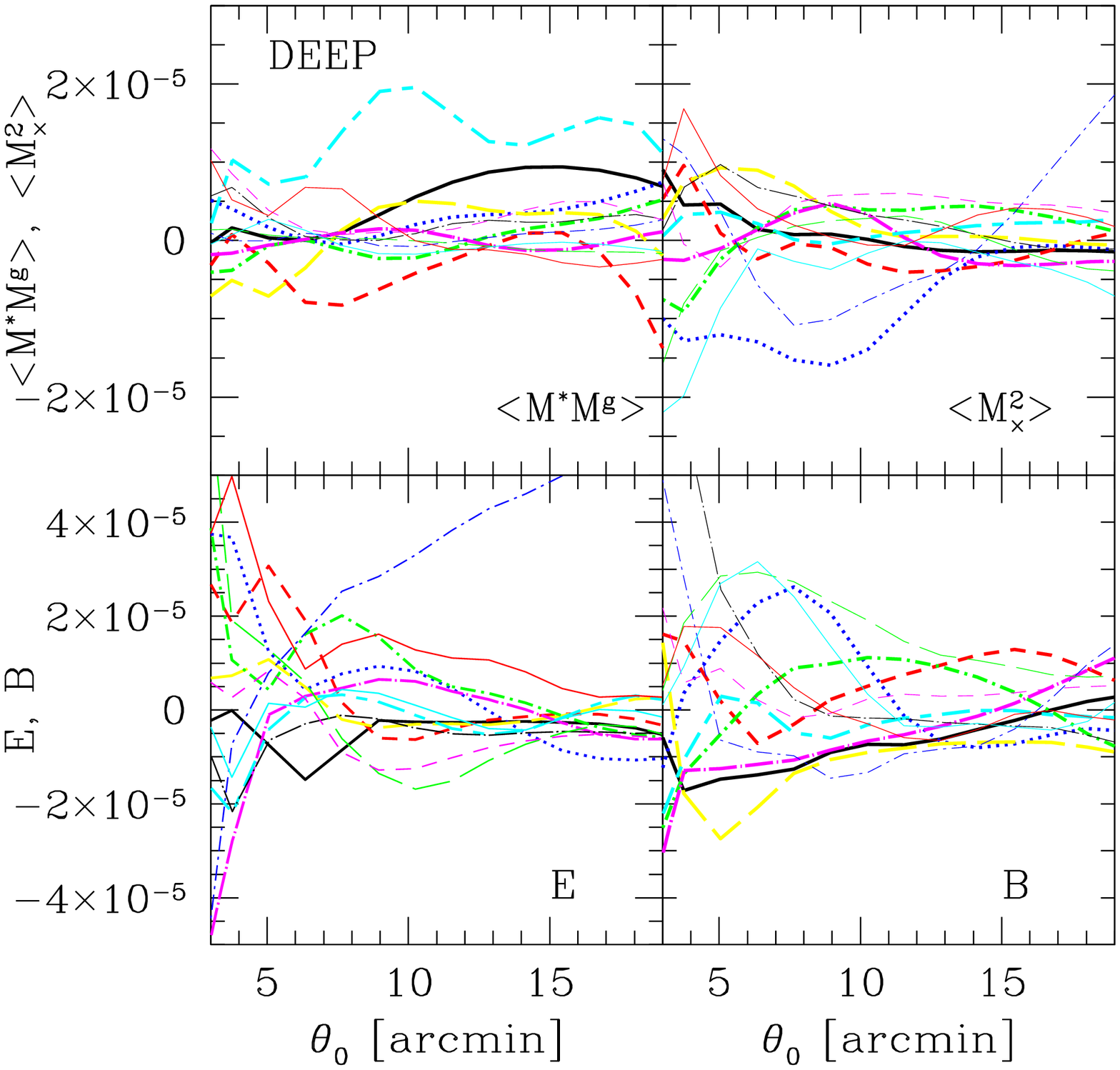}
\includegraphics[scale=0.38]{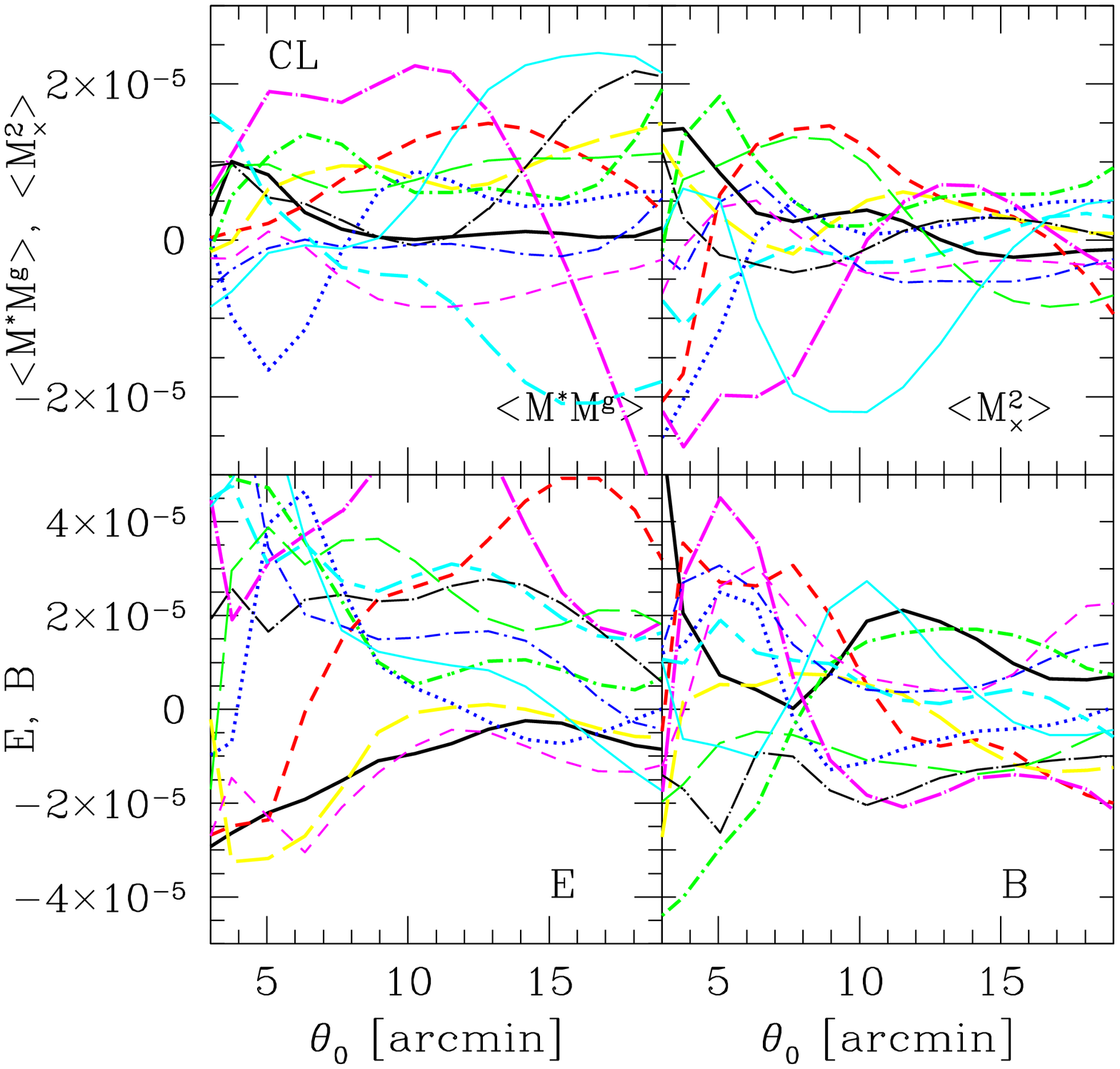}
\includegraphics[scale=0.38]{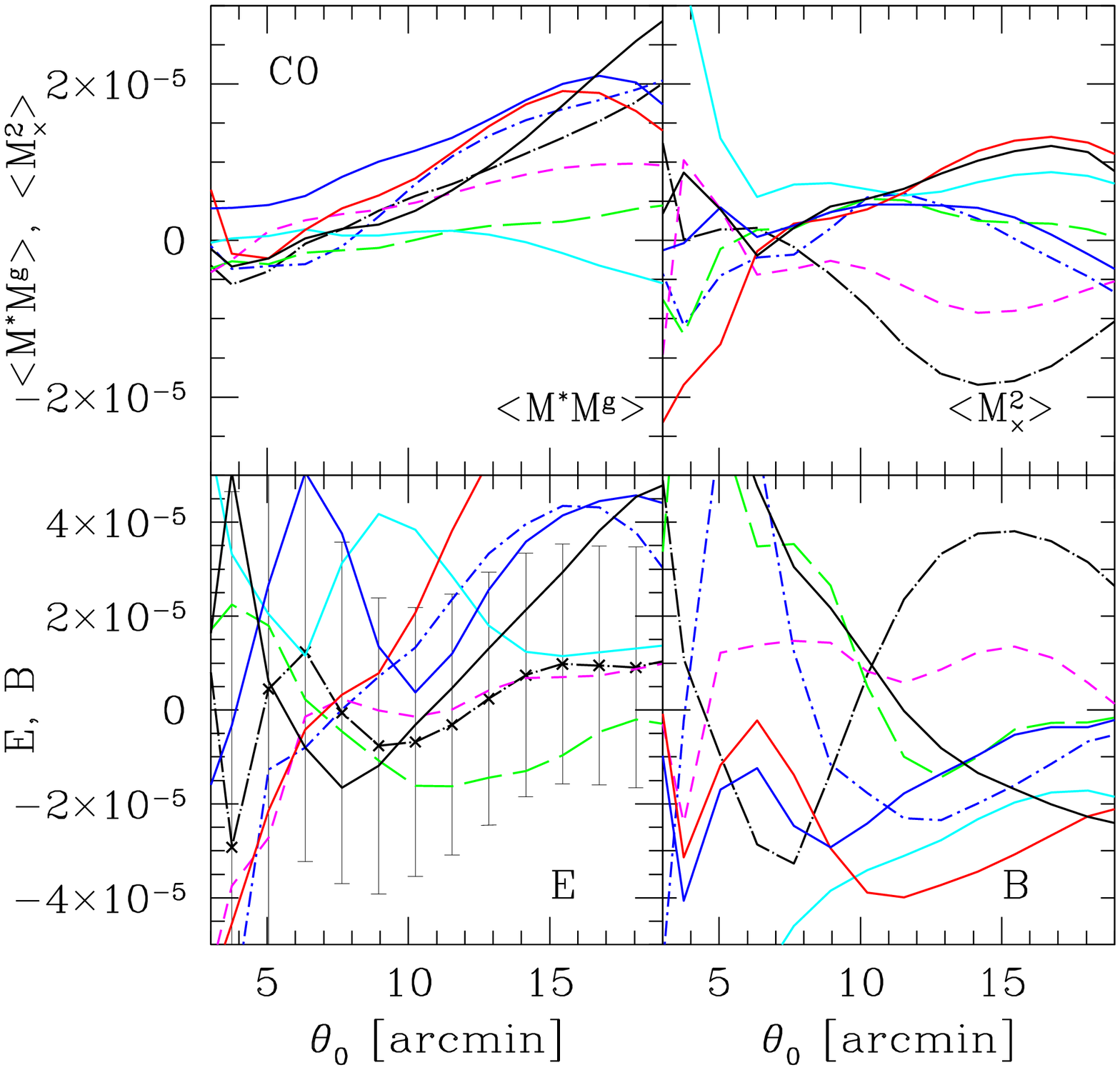}
\includegraphics[scale=0.38]{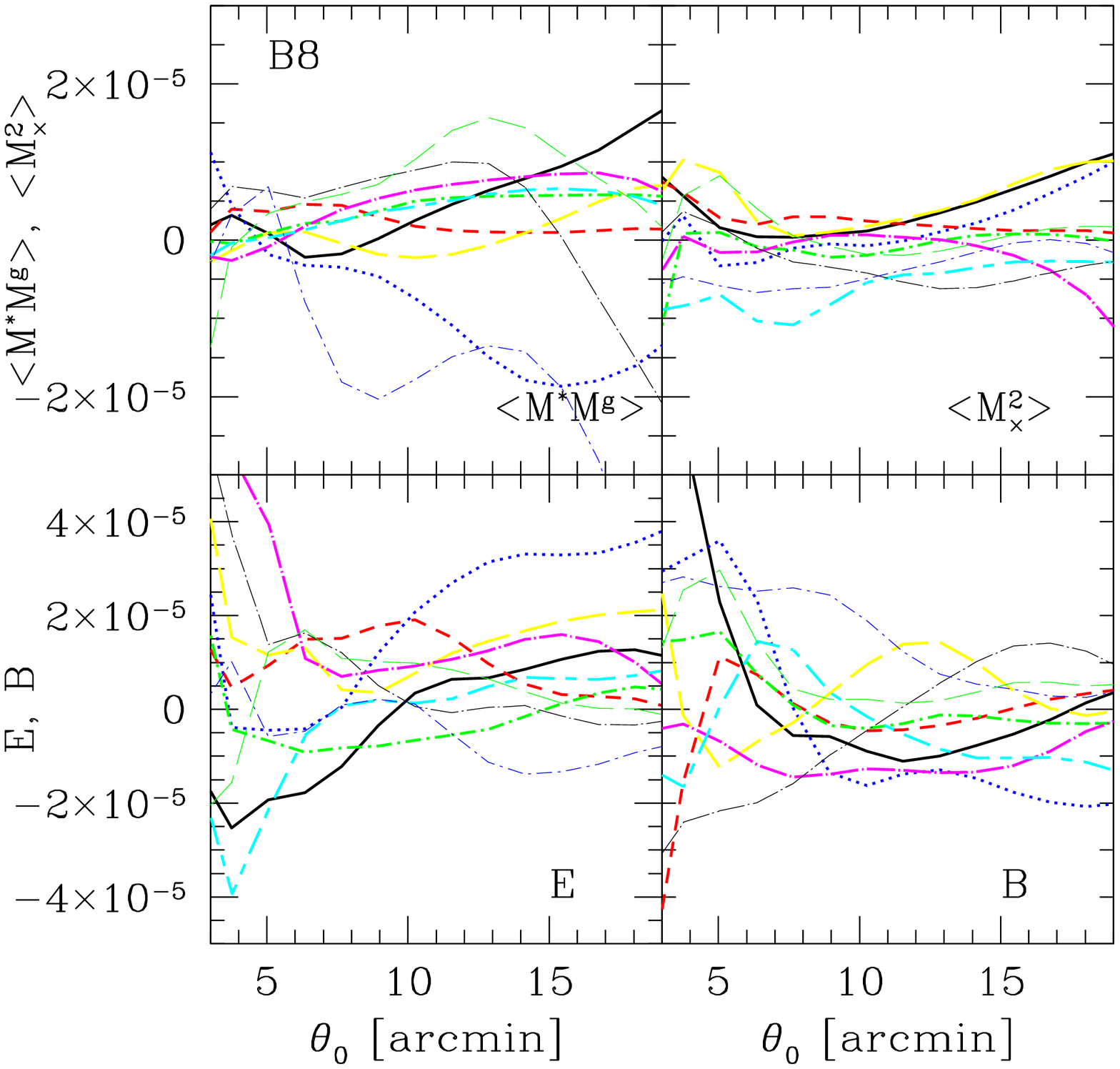}
\caption{
In each diagram various $M_{\rm ap}$ statistics are displayed for each field:
cross-correlation between uncorrected stars and anisotropy corrected galaxies, $\langle M^*M^{\rm g}\rangle$,
cross-correlation between the tangential component and the cross component of the shear, $M_\times$,
E and B-mode. 
The name codes are given in table \ref{SumTab}.
For two fields we plotted the statistical errors: FDF (upper left panel) and C04p3 (lower left panel).
The type of each line is the same as in Fig. \ref{fig:weight}. 
}
\label{fig:singlefields}
\end{figure*}
%
%
%
\end{document}